 \documentclass[final,3p,times]{elsarticle1}

\usepackage{amssymb,amsmath,verbatim,dsfont,url,float,hyperref}

\biboptions{sort&compress}

\newcommand{\str}{\text{\,str\,}}
\newcommand{\tr}{\text{\,tr\,}}
\newcommand{\sdet}{\text{sdet\,}}
\newcommand{\diag}{\text{diag\,}}
\newcommand{\dd}{\mathrm{d}}
\newcommand{\1}{\mathds{1}}
\newcommand{\bSg}{\boldsymbol{\Sigma}}

\begin{document}

\begin{frontmatter}

\title{Distributions of Off-Diagonal Scattering Matrix Elements: Exact Results}


\author{A. Nock\footnote{Present address: \textit{School of Mathematical Sciences, Queen Mary University of London, London E1 4NS, United Kingdom}}}
\ead{a.nock@qmul.ac.uk}
\author{S. Kumar\footnote{Present address: \textit{Department of Physics, Shiv Nadar University, Gautam Budh Nagar, Uttar Pradesh - 203207, India}}}
\ead{skumar.physics@gmail.com}
\author{H.-J. Sommers}
\ead{h.j.sommers@uni-due.de}
\author{T. Guhr}
\ead{thomas.guhr@uni-due.de}
\address{Fakult\"at f\"ur Physik, Universit\"at Duisburg-Essen, Lotharstra\ss{}e 1, 47048 Duisburg, Germany}

\begin{abstract}
Scattering is a ubiquitous phenomenon which is observed in a variety of physical systems which span a wide range of length scales. The scattering matrix
 is the key quantity which provides a complete description of the scattering process. The universal features of scattering in chaotic systems is most generally 
 modeled by the Heidelberg approach which introduces stochasticity to the scattering matrix at the level of the Hamiltonian describing the scattering center. 
 The statistics of the scattering matrix is obtained by averaging over the ensemble of random Hamiltonians of appropriate symmetry. We derive exact results 
 for the distributions of the real and imaginary parts of the off-diagonal scattering matrix elements applicable to orthogonally-invariant and unitarily-invariant 
 Hamiltonians, thereby solving a long standing problem. 
 \end{abstract}


\end{frontmatter}



\section{Introduction}

Scattering is a truly fundamental issue in physics~\cite{Scattering,FKS2005}. A major part of our information about quantum systems stems from scattering
 experiments. Rutherford's gold-foil experiment~\cite{RGM1911} is a classic example which led us towards the understanding of the atomic structure. 
 Even in modern times, powerful particle accelerators rely on scattering experiments to probe deeper and deeper into the structure of matter. Moreover, 
 scattering plays a crucial role in classical wave systems as well and one can often relate the relevant observables to the scattering parameters. Along 
 with the atomic nuclei~\cite{FPW1954,R1991,BLR1993,DPS1996,MRW2010}, atoms~\cite{TNC1989, GDG1993, DZD1995,MG1994} and 
 molecules~\cite{Gaspard1993,SKSD1995,LS1993}, some of the other important examples where scattering phenomena have been of considerable interest are 
 mesoscopic ballistic devices ~\cite{BM1994,Stone1995,SS1988,IWZ1990,Altland1991,Zirnbauer1992,MMZ1994,PWZLW1995,MB1995,BM1996,Beenakker1997,MK2004}, microwave cavities~\cite{SS1990,Sridhar1991,
 SSS1995,DSF1990,SAOO1995,ABGHPRS1993,AGHHLRSW1995,SKBLS2003,KMMS2005,HZAOA2005,HHZAOA2006,KHMS2008,D2009,
 YHBAOA2010,D2010a,D2010b}, irregular graphs~\cite{HTBKS2005,LHBSS2008}, quantum graphs~\cite{KS2000,GA2004,PW2013}, elastomechanical
  billiards~\cite{EGLNO1996,KSW2005,AEOGS2010}, wireless communication~\cite{YAOA2012,YOAA2012,BC2001} etc. 

The scattering process can be completely described in terms of the scattering matrix ($S$ matrix). It relates the asymptotic initial and final Hilbert spaces spanned
 by a quantum system undergoing the scattering process. In simple words, it relates the incoming and outgoing
 waves. In a quantum mechanical context these are the wave functions, i.e. the probability amplitudes. However, in classical systems, 
 the waves are the displacement vectors in elastomechanical systems or the electromagnetic field in microwave cavities. 
The flux conservation requirement constrains the $S$ matrix to be unitary, i.e., $SS^\dag=S^\dag S=\mathds{1}$. As a consequence of the complicated 
 dependence on the parameters of the incoming waves and the scattering center, scattering is quite often of chaotic nature. Accordingly one needs a statistical 
 description of the scattering phenomenon and hence of the $S$ matrix, i.e., to describe the $S$ matrix and related observables in terms of correlations functions 
 and distributions. Two standard approaches in this direction are the semiclassical approach~\cite{BS1989,Smilansky1990,DS1992,Richter2000} and the 
 stochastic approach~\cite{AWM1975,Weidenmuller1992, Mello1993,MPS1985}. The former relies on representing the $S$-matrix elements in terms of a sum 
 over the classical periodic orbits, starting with the genuine microscopic Hamiltonian representing the system. The latter, in contrast, relies on introducing 
 stochasticity to the scattering matrix or to the Hamiltonian describing the scattering center. Both of these have their advantages and drawbacks. For instance,
  the semiclassical approach suffers the restriction caused by an exponential proliferation of classical periodic unstable trajectories. It is further constrained by 
  the formal condition $\hbar \rightarrow 0$ which demands that the number of open channels be large and therefore does not cover all interesting cases. The 
  stochastic approach, on the other hand, gets restricted by the very nature of the stochastic modeling. Moreover, in this case, one can expect only to explore the
   universal aspects, leaving aside the system specific properties. The comparison between these two approaches has been discussed in detail in~\cite{LW1991}. 

As indicated above, within the stochastic approach, one can pursue one of the following two routes. In the first one, the $S$ matrix itself is regarded a stochastic 
quantity and is described by the Poisson kernel. Its derivation is based on imposing minimal information content along with the necessary conditions like unitarity, 
analyticity etc. This route was pioneered by Mello and coworkers and is often referred to as the Mexico approach~\cite{Mello1993,MPS1985}. The second path 
relies on introducing the stochasticity at the level of the Hamiltonian describing the scattering center. For this, one employs the random matrix universality 
conjecture and models the system Hamiltonian by one of the appropriate random matrix ensembles~\cite{BFFMPW1981,Mehta2004,GMW1998}. This path
 was laid by Weidenm\"{u}ller and coworkers~\cite{AWM1975} and is referred to as the Heidelberg approach. Even though these two stochastic approaches 
 appear very different in their formulation, they describe precisely the same quantity, the $S$ matrix. Naturally, one would expect that these two routes are 
 equivalent. Indeed it was shown by Brouwer that the Poisson kernel can be derived using the Heidelberg approach by modeling the scattering-center Hamiltonian
 by a Lorentzian (or Cauchy) ensemble of random matrices~\cite{Brouwer1995}. Since the universal properties depend only on the invariance properties of the 
 underlying Hamiltonian~\cite{BFFMPW1981,Mehta2004,GMW1998}, his result established the equivalence between the two approaches. Furthermore, very
  recently Fyodorov {\it et al.} have demonstrated this equivalence for a broad class of unitary-invariant ensembles of random matrices~\cite{FKN2013}.

In their pioneering work Verbaarschot {\it et al.}~\cite{VWZ1985} calculated the two-point energy correlation functions by implementing the supersymmetry
 technique~\cite{Efetov1983,Berezin1987,Efetov1997,Guhr2010} within the Heidelberg approach. Their result established the universality of the $S$-matrix 
 fluctuation properties in chaotic scattering. Further progress in characterizing the $S$-matrix fluctuations was made in~\cite{DB1988,DB1989} where the authors
 derived up to the fourth-moment. In Refs.~\cite{RFW2003,RFW2004} a related problem of statistics of transmitted power in complex disordered 
 and ray-chaotic structures was solved. The Landauer-B\"{u}ttiker formalism~\cite{Beenakker1997,MK2004,SS1988} gives the quantum conductance of mesoscopic systems 
 (quantum dots and quantum wires) in terms of the scattering matrix elements. In these systems the Heidelberg approach has been used to calculate 
 the average and variance of conductance in Refs.~\cite{IWZ1990,Altland1991,Zirnbauer1992,MMZ1994,PWZLW1995}. 
 These results served as important steps in our understanding of the nature of scattering in chaotic systems. However, 
 a more stringent investigation of the universality at the level of individual $S$-matrix elements requires information beyond that of a first few
 moments~\cite{SKBLS2003,KMMS2005,HZAOA2005, HHZAOA2006,KHMS2008,D2009,YHBAOA2010,D2010a,D2010b}. A complete description is provided
 only by the full distributions which is equivalent to having the knowledge of all the moments. In the limit of a large number of open channels and a vanishing 
 average $S$ matrix, or equivalently, in the Ericson regime of strongly overlapping resonances~\cite{MRW2010,AWM1975}, the real and the imaginary parts 
 of the $S$-matrix elements exhibit Gaussian behavior. However, outside this regime the unitarity of the $S$ matrix results in significant deviations from the Gaussian 
 distribution~\cite{MRW2010,DB1988,DB1989,T1975,RSW1975}. The available moments up to the fourth are insufficient to determine the exact behavior of 
 these distributions in a general case. A significant progress in characterizing the behavior of diagonal $S$-matrix elements in the general case was made
 in~\cite{FSS2005} where the authors succeeded in deriving the full distributions. The off-diagonal elements, however, could not be tackled by the well 
 established methods and the problem of finding their distributions remained unsolved till very recently~\cite{KNSGDMRS2013}. Here, we derive an exact
 solution to this problem and present results which are valid in all regimes. We gave a brief presentation of the results in Ref.~\cite{KNSGDMRS2013}. In 
 the present work we provide a full-fledged derivation with all the details, as well as some more new results concerning the statistics of off-diagonal $S$-matrix
  elements. Our approach is based on a novel route to the nonlinear sigma model which involves obtaining the {\it characteristic functions} associated with the 
  distributions. This is different from the usual approach where one formulates an appropriate generating function for the $S$-matrix correlations. By contrast, the 
  characteristic function is the moment generating function.
 
The paper is arranged as follows. In Section \ref{secSM} we set up the model for the scattering process using the Hamiltonian formulation and implement the 
Heidelberg approach to introduce stochasticity. In Section \ref{secDCF} we define the quantities to be calculated, viz., the probability distributions for off-diagonal 
matrix elements and the associated characteristic functions. Section \ref{secUE} deals with the exact results for unitarily-invariant Hamiltonians, which apply to 
systems with broken time-reversal invariance. Section \ref{secOE} gives the exact results for orthogonally-invariant Hamiltonians, which apply to ``spinless" 
time-reversal invariant systems. We conclude in Section \ref{secConc} with a brief discussion. In the appendices, we collect some of the derivations.

\section{Scattering Matrix}
\label{secSM}

In the generic setting of the scattering problem the scattering event is assumed to take place inside only a certain part of the available space. Outside this 
``interaction region'' the fragments exhibit a free motion which is characterized, along with the energy $E$, by a set of quantum numbers. The states 
corresponding to these quantum numbers represent the states in which the fragments exist asymptotically before or after the scattering event and are referred
to as {\it channels of reaction}.

We associate with the compact interaction region a discrete set of orthogonal states $|n\rangle; n=1,2,...,N$, which represent the bound states of the Hamiltonian
 $H$ describing the ``closed'' chaotic system. Moreover, we assume that at given energy $E$ there are exactly $M$ open channels, described by a continuous 
 set of functions $|c,E\rangle; c=1, . . . ,M$ satisfying the orthogonality condition $\langle a,E_1|b,E_2\rangle=\delta_{ab}\,\delta(E_1-E_2)$. The full Hamiltonian 
 $\mathcal{H}$ for the system can therefore be written as
\begin{equation}
\label{Hamilt}
 \mathcal{H}=\mathcal{H}_0+\mathcal{V}.
\end{equation}
Here $\mathcal{H}_0$ describes the part of the Hamiltonian which is present without any interaction between the internal states of the system Hamiltonian $H$
 and states of the open channels, viz.,
\begin{equation}
\label{H0}
 \mathcal{H}_0=\sum_{l,m} |l\rangle H_{lm} \langle m|+\sum_{c=1}^M \int_{\epsilon_c}^\infty dE |c,E\rangle E \langle c, E|,
\end{equation}
and $\mathcal{V}$ represents the interaction part,
\begin{equation}
 \mathcal{V}=\sum_{l,c} \int_{\epsilon_c}^\infty dE\left(|l\rangle~ (W_{c})_l~\langle c,E| + \text{herm. conj.}\right).
\end{equation}
Here $\epsilon_c$ represents the threshold energy in a given channel $c$, and thus integrals in the above two equations run over the energy region where the 
channel $c$ is open. $W_c\, (c=1,...,M)$ are the $N$-dimensional coupling vectors which encode the information about the interaction. In Eq. \eqref{H0} any 
direct interaction between channels has been neglected for simplicity, thereby rendering the second term diagonal in $c$. Furthermore, the dependence of 
the coupling vectors on energy has also been ignored as we are interested in a situation where the mean level spacing between the resonances is very 
small compared to the mean level spacing between the channel thresholds.

Under some reasonable assumptions the $S$-matrix elements can be obtained in terms of the Hamiltonian $H$ and the coupling vectors $W_c$ as~\cite{MW1969,FS1997}
\begin{equation}
\label{Sab}
 S_{ab}(E)=\delta_{ab}-i 2\pi W_a^\dag G(E) W_b,
\end{equation}
where the inverse of the resolvent $G(E)$ is given by
\begin{equation}
\label{G}
G^{-1}(E)=E\mathds{1}_N-H+i\pi\displaystyle\sum_{c=1}^M W_c W_c^\dag.
\end{equation} 
The above $S$-matrix ansatz provides the most general description of any scattering process in which an interaction zone and scattering channels can be identified. 
For the characterization of the $S$ matrix one has to specify properties of the coupling vectors $W_c$. A convenient choice corresponds to the case when the 
average $S$ matrix is diagonal, viz., $\overline{S}_{ab}=\overline{S}_{aa} \delta_{ab}$ \cite{VWZ1985,LW1991}. In this case the coupling vectors $W_c$ can be 
chosen to obey the following orthogonality relation \cite{VWZ1985}: 
\begin{equation}
\label{Wortho}
W_c^\dag W_d=\frac{\gamma_c}{\pi} \delta_{cd}.
\end{equation}
This choice corresponds to the absence of any direct coupling between the channels \cite{VWZ1985,LW1991}. An alternative choice which also fulfills the condition
 of $\overline{S}$ being diagonal is of considering the elements of $W_c$ as zero-mean independent Gaussians with variances proportional to $\gamma_c$. 
 It turns out that these two choices are equivalent as long as $M\ll N$~\cite{LSSS1995a, LSSS1995b}, which is exactly the case we are interested in. We consider
 the former choice in our calculations , i.e., Eq. \eqref{Wortho}. We would like to remark here that for the case of a non-diagonal average $S$ matrix, a unitary
 transformation $U$ can always be found such that $USU^\dag$ is diagonal on average and has the same fluctuation properties as the $S$ matrix without direct 
 reactions~\cite{VWZ1985, EW1973}. Thus it suffices to consider a case which omits direct reactions. 

We now evoke the random matrix universality conjecture, according to which the universal and generic properties of chaotic systems can be extracted by modeling
 the underlying Hamiltonian (or its analogue) by an ensemble of random matrices of appropriate symmetry class. We consider here the Gaussian ensemble of 
 random matrices to model the interaction-region Hamiltonian $H$. This particular choice of distribution is only for calculation convenience since it is known that 
 for the universal properties, as long as one takes into account the proper invariance properties of the Hamiltonian to be modeled, the choice of distribution is
  immaterial. See for example~\cite{MZ2010} where the authors calculate the two-point correlation function by considering an arbitrary $U(N)$ invariant Hamiltonian. 

Depending on whether the system is time-reversal invariant or noninvariant, $H$ is chosen to belong to the Gaussian Orthogonal Ensemble (GOE) or the Gaussian
Unitary Ensemble (GUE)~\cite{BFFMPW1981,Mehta2004,GMW1998}. These two ensembles are designated by the Dyson index $\beta$ and have the following 
probability distribution associated with them:
\begin{equation}
\label{Hdist}
 \mathcal{P}(H)\propto \exp\left(-\frac{\beta N}{4v^2}\text{tr}H^2\right).
\end{equation}
The GOE and GUE are described respectively by $\beta=1$ and $\beta=2$. $N$ in the above equation represents the dimensionality of $H$ which is 
essentially the number of bound states, and $v^2$ is a free parameter which can be chosen to fix the energy scale. For $\beta=1$, $H$ is a real-symmetric matrix
 and has $N(N+1)/2$ independent parameters. On the other hand for $\beta=2$, $H$ is Hermitian and involves $N^2$ independent parameters. For 
 $N\rightarrow \infty$ we obtain from Eq.~\eqref{Hdist}, for both values of $\beta$, the density of eigenvalues as the Wigner semicircle~\cite{Mehta2004},
\begin{equation}
\label{Wsc}
\rho(E)=\frac{1}{2\pi v^2}\sqrt{4v^2-E^2}. 
\end{equation}
The level density is $N \rho(E)$, and consequently the mean level spacing is $\Delta_m=1/(N\rho(E))$, which for large $N$ behaves as $1/N$ in the bulk of the spectrum.

\section{Distributions and Characteristic Functions}
\label{secDCF}

We are interested in the statistics of the off-diagonal elements of the $S$ matrix. The off-diagonal $S$-matrix elements relate the amplitudes in different channels. 
Thus their statistical information is as important, if not more, as that of the diagonal $S$-matrix elements which relate the amplitudes within the same channel. The
 $S$-matrix elements being complex quantities, we need to investigate the behavior of their real and imaginary parts or equivalently that of their moduli and phases. 
 We consider here the distributions of the real and imaginary parts and deal with them simultaneously. We introduce the notation $\wp_s(S_{ab})$, with $s=1,2$ 
 to refer to the real and imaginary parts of $S_{ab}$ respectively. For the off-diagonal case, by setting $a\neq b$ we obtain from Eq. \eqref{Sab},
\begin{eqnarray}
\wp_s(S_{ab})=\pi \big((-i)^sW_a^\dag G W_b +i^s W_b^\dag G^\dag W_a\big),~~~ s=1,2.
\end{eqnarray}
To find the corresponding distributions, $P_s(x_s)$, we need to perform the following ensemble average:
\begin{equation}
\label{Ps1}
P_s(x_s)=\int d[H] \mathcal{P}(H)\delta\left(x_s-\wp_s(S_{ab})\right).
\end{equation}
Here the volume element $d[H]$ represents the flat measure involving the product of the differentials of all independent variables occurring within $H$. As 
mentioned in the introduction, we analyze the corresponding characteristic function, viz.
\begin{equation}
\label{Rsk}
R_s(k)=\int d[H]\mathcal{P}(H) \exp(-i k \wp_s(S_{ab})).
\end{equation}
The characteristic function also serves as the moment generating function, i.e., all the moments of the real and imaginary parts of the $S$-matrix elements can 
be obtained by expanding $R_s(k)$ in powers of $k$. The expression for $P_s(x_s)$ can be retrieved from the Fourier transform of $R_s(k)$ as
\begin{equation}
\label{Ps2}
P_s(x_s)=\frac{1}{2\pi}\int_{-\infty}^\infty dk  R_s(k) \exp(i k x_s).
\end{equation}
Thus our strategy is to calculate the characteristic function first and then obtain the distribution from it by taking the Fourier transform.

We introduce a $2N$-component vector $W$ involving the coupling vectors $W_a, W_b$, and a $2N\times2N$-dimensional matrix $A_s$ composed of the 
resolvent $G$ as
\begin{equation}
W=\begin{bmatrix}W_a \\ W_b \end{bmatrix}_{2N}, ~~~A_s=\begin{bmatrix} 0 & (-i)^s G \\ i^s G^\dag & 0 \end{bmatrix}_{2N\times 2N}.
\end{equation}
The characteristic function $R_s(k)$ can be expressed in terms of these quantities as
\begin{equation}
\label{RkWdAW}
R_s(k)=\int d[H] \mathcal{P}(H) \exp(-i k \pi W^\dag A_s W).
\end{equation}
To evaluate $R_s(k)$ we need to integrate over the ensemble of $H$-matrices defined by Eq.~\eqref{Hdist}. In general this is a nontrivial task, more so when 
the quantity to be averaged does not respect the invariance properties of $H$, which is the case here. Further complications are caused here because of the
 extremely convoluted dependence of $H$ in the exponent in Eq.~\eqref{RkWdAW} -- it appears in the denominator of the resolvent $G$ contained in the
 matrix $A_s$. To overcome this problem we seek some trick which will invert the $G$ in Eq.~\eqref{RkWdAW}, thereby rendering the exponent linear in
 $H$. As we will see below, the supersymmetry formalism provides exactly such a solution~\cite{Efetov1983,Berezin1987,Efetov1997,Guhr2010}.

We introduce a $2N$-dimensional complex vector $z^T=[z_a^T, z_b^T]=[z_{a1},...,z_{aN},z_{b1},...,z_{bN}]$ consisting of commuting (Bosonic) variables. 
Similarly we introduce a $2N$-dimensional vector $\zeta^T=[\zeta_a^T, \zeta_b^T]=[\zeta_{a1},...,\zeta_{aN},\zeta_{b1},...,\zeta_{bN}]$ consisting of 
anticommuting (Fermionic or Grassmann) variables. We note that the indices $a,b$ in these vectors are just dummy indices and do not have any direct 
dependence on the values of the indices signifying the $S$-matrix element. We now consider the following multivariate Gaussian integral results for commuting
 and anticommuting variables:
\begin{equation}
\label{gintc}
\int d[z] \exp\big[i (z^\dag {\bf a}\, z+{\bf b}^\dag z+z^\dag {\bf c})\big]={\det}^{-1}\left(\frac{{\bf a}}{2\pi i}\right) \exp(-i {\bf b}^\dag {\bf a}^{-1} {\bf c}), 
\end{equation}
\begin{equation}
\label{gintac}
\int d[\zeta] \exp\big[i (\zeta^\dag {\bf a} \,\zeta+\boldsymbol{\mu}^\dag \zeta+\zeta^\dag \boldsymbol{\nu})\big]=\det\left(\frac{{\bf a}}{2\pi i}\right)
\exp(-i \boldsymbol{\mu}^\dag {\bf a}^{-1}\boldsymbol{\nu}). 
\end{equation}
In Eqs. (\ref{gintc}) and (\ref{gintac}), $\bf{a}$ is an arbitrary normal matrix with complex entries. {\bf b}, {\bf c} in Eq. \eqref{gintc} are vectors consisting of commuting 
entries, while $\boldsymbol{\mu}, \boldsymbol{\nu}$ in Eq. \eqref{gintac} are vectors having anticommuting entries. The volume elements $d[z]$ and $d[\zeta]$
 in the above equation are given by $d[z]=\prod_{j=1}^N dz_{aj}^*dz_{aj}\, dz_{bj}^*dz_{bj}$ and $d[\zeta]=\prod_{j=1}^N d\zeta_{aj}^*d\zeta_{aj}\, d\zeta_{bj}^*d\zeta_{bj}$.
  The above two Gaussian-integral identities enable us to recast the characteristic function, Eq. \eqref{RkWdAW}, in the following form:
\begin{eqnarray}
\label{Rkcac}
R_s(k)=\int d[z]\exp\left[\frac{i}{2}(z^\dag W+W^\dag z)\right] \int d[\zeta]\int d[H] 
\mathcal{P}(H)\exp\left[\frac{i}{4\pi k} (z^\dag A_s^{-1} z+\zeta^\dag A_s^{-1} \zeta)\right].
\end{eqnarray}
Eq.~\eqref{Rkcac} can also be expressed in  terms of an integral over a $4N$-dimensional supervector $\Psi^T=[z^T,\zeta^T]$ as:
\begin{equation}
\label{RkA}
 R_s(k)=\int d[\Psi] \exp\left[\frac{i}{2}(\mathbf{W}^\dag \Psi+\Psi^\dag \mathbf{W})\right]\int d[H] 
\mathcal{P}(H)\exp\left(\frac{i}{4\pi k} \Psi^\dag \mathbf{A}_s^{-1} \Psi\right).
\end{equation}
Here $\mathbf{A}_s^{-1}=\mathds{1}_2\otimes A_s^{-1}$ and $\mathbf{W}^\dag=[W^\dag,0]$ are $4N$-dimensional square-matrix and vector respectively. An 
ensemble average over an exponential of a bilinear form involving supervectors and a matrix, as in the equation above, is common in supersymmetry calculations. 
However, there is a difference here: $\mathbf{A}_s^{-1}$ is not in block-diagonal form. If we carry out the ensemble average using this form of  $\mathbf{A}_s^{-1}$
 it will result in problems incorporating the correct symmetry properties in the supermatrix which has  to be introduced later. To resolve this problem we employ the
 trick of carrying out certain transformations in $z$ and $\zeta$, while leaving $z^\dag$ and $\zeta^\dag$ as they are. This is allowed since $z\, (\text{resp. }\zeta)$
 and $z^\dag (\zeta^\dag)$, being complex quantities, admit independent transformations.

To proceed further from this point, we have to take into account the appropriate symmetry of $H$, i.e., whether it belongs to the GOE or GUE.

\section{Unitarily Invariant Hamiltonians ($\beta=2$)}
\label{secUE}

We consider in this section the case when $H$ is modeled by the GUE, and thus is applicable to systems with completely broken time-reversal symmetry. In this case
 the Hamiltonian $H$ is complex-Hermitian \cite{BFFMPW1981,Mehta2004,GMW1998}. The route to the final results will consist of three steps: (i) mapping the above
  result to a matrix integral in superspace, (ii) applying the large $N$-limit and obtaining the nonlinear $\sigma$-model, and finally (iii) reducing the result to integrals
   over the radial coordinates.
 
 \subsection{Mapping to a matrix integral in superspace}
 
 As mentioned in the previous section we want to bring $\mathbf{A}_s^{-1}$ in Eq.~\eqref{RkA} into a block-diagonal form. To accomplish this we consider the 
 following transformations in the vectors: 
\begin{equation}
z\rightarrow \Xi^{+} z,~~ z^\dag\rightarrow z^\dag, 
~~\zeta\rightarrow  \Xi^{-} \zeta,~~\zeta^\dag\rightarrow\zeta^\dag,
\end{equation}
where
\begin{equation}
\label{trans}
 \Xi^{\pm}=\begin{bmatrix} 0 & \pm(-i)^s\mathds{1}_N \\ -i^s\mathds{1}_N & 0 \end{bmatrix}.
\end{equation}
The different transformations for $z$ and $\zeta$ ensure proper symmetry and convergence properties of the supermatrix $\sigma$ when we map the problem
 to a matrix integral in superspace. The Jacobian factor arising from the above transformations is $(-1)^N$. We therefore obtain
\begin{equation}
\label{RkU}
 R_s(k)=(-1)^N\int d[\Psi] \exp\left[\frac{i}{2}(\mathbf{U}_s^\dag \Psi+\Psi^\dag \mathbf{W})\right]\int d[H] 
\mathcal{P}(H)\exp\left(\frac{i}{4\pi k} \Psi^\dag \boldsymbol{\mathcal{A}}^{-1} \Psi\right),
\end{equation}
where $\mathbf{U}_s^\dag=[-i^s W_b^\dag,(-i)^s W_a^\dag,0,0]$ and results from $\mathbf{W}^\dag$ of Eq.~\eqref{RkA} because of the rotation of the supervector 
$\Psi$. The new $4N$-dimensional matrix $\boldsymbol{\mathcal{A}}^{-1}=\text{diag}[-(G^{-1})^\dag,G^{-1},-(G^{-1})^\dag,-G^{-1}]$ in the above equation is block
 diagonal as desired. 

We now examine the  $H$-dependent part in the exponent in Eq. \eqref{RkU}. It possesses the bilinear form, 
$z_a^\dag H z_a-z_b^\dag H z_b+\zeta_a^\dag H \zeta_a+\zeta_b^\dag H \zeta_b$, which can also be written as $\tr HD$, where
\begin{equation}
D=z_a z_a^\dag-z_b z_b^\dag-\zeta_a \zeta_a^\dag-\zeta_b \zeta_b^\dag.
\end{equation}
The Hermiticity of the matrix $D$ is evident. The GUE averaging in Eq. \eqref{RkU}  therefore amounts to performing the following integral:
\begin{equation}
\label{ens_av}
\int \,d[H] \mathcal{P}(H)  \exp\left(\frac{i}{4\pi k} \tr HD\right)
=\exp\left(-\frac{1}{4 r}\,\text{tr}\,D^2\right),
\end{equation}
where the variable $r$ incorporates parameters of the problem as 
\begin{equation}
r=\frac{8\pi^2k^2 N}{v^2}.
\end{equation}
The expression on the right hand side of Eq. \eqref{ens_av} can also be written in terms of the supertrace (str)~\cite{Efetov1997,Guhr2010} involving a 4-dimensional
 supermatrix $\mathcal{B}$ having elements
\begin{equation}
\label{Bmn}
\mathcal{B}_{m n}= \sum_{j=1}^N (\Psi_m)_j (\Psi_n^\dag)_j,
\end{equation}
where $m,n=1,...,4$ and $\Psi_1\equiv z_a,\Psi_2\equiv z_b,\Psi_3\equiv\zeta_a$ and $\Psi_4\equiv\zeta_b$. The Boson-Boson and Fermion-Fermion blocks of
 the supermatrix $\mathcal{B}$ are Hermitian, while the other two blocks are adjoints of each other, and therefore $\mathcal{B}$ is Hermitian itself. We have
\begin{equation}
\label{trstr}
\exp\left(-\frac{1}{4 r}\,\text{tr}\,D^2\right)=\exp\left[-\frac{1}{4 r}\str(K^{1/2}\mathcal{B} K^{1/2})^2\right] 
\end{equation}
with  $K=\text{diag}(1,-1,1,1)$. 
Eq. \eqref{trstr} demonstrates the duality between the ordinary space and the superspace~\cite{Guhr2010}. The characteristic function can now be written as
\begin{equation}
\label{RkLBL_UE}
R_s(k)=(-1)^N\int d[\Psi] \exp\left[-\frac{1}{4r}\str(K^{1/2}\mathcal{B} K^{1/2})^2\right]\exp\left[\frac{i}{4\pi k} \Psi^\dag \boldsymbol{\mathcal{A}}_0^{-1} \Psi
+\frac{i}{2}(\mathbf{U}_s^\dag \Psi+\Psi^\dag \mathbf{W})\right]. 
\end{equation}
We have introduced here $\boldsymbol{\mathcal{A}}_0^{-1}$ which is the $H$-independent part of $\boldsymbol{\mathcal{A}}^{-1}$, viz.
\begin{equation}
\label{A0_UE}
\boldsymbol{\mathcal{A}}_0^{-1}=-\mathbf{K}^{1/2}\left(E\mathds{1}_{4N}-L\otimes i\pi\sum_{c=1}^M W_c W_c^\dag\right)\mathbf{K}^{1/2},
\end{equation}
with $\mathbf{K}=K\otimes \mathds{1}_N$ and $L=\diag(1,-1,1,-1)$.
We now use the Hubbard-Stratonovich identity~\cite{VWZ1985,Efetov1983,Efetov1997,Guhr2010}, 
\begin{equation}
\label{HSid}
\exp\left[-\frac{1}{4r}\str(K^{1/2}\mathcal{B} K^{1/2})^2\right]=\int d[\sigma] 
\exp\left(-r\str \sigma^2+ i\str \sigma K^{1/2}\mathcal{B} K^{1/2}\right),
\end{equation}
and map the problem to a matrix integral in superspace by introducing a 4-dimensional supermatrix $\sigma$ having same symmetry as
$\mathcal{B}$. Also, observing that 
\begin{equation}
i\str \sigma K^{1/2}\mathcal{B} K^{1/2}=i\,\Psi^\dag \mathbf{K}^{1/2}(\sigma \otimes \mathds{1}_N)\mathbf{K}^{1/2}\Psi
\end{equation}
 we arrive at
\begin{eqnarray}
\label{susym}
R_s(k)=(-1)^N\int d[\sigma]\exp(-r\str\sigma^2)\int d[\Psi] 
\exp\Big[i \Psi^\dag \mathbf{K}^{1/2}\boldsymbol{\Sigma}\,\mathbf{K}^{1/2} \Psi
+\frac{i}{2}(\mathbf{U}_s^\dag \Psi+\Psi^\dag \mathbf{W})\Big],  
\end{eqnarray}
where
\begin{equation}
\boldsymbol{\Sigma}=\sigma_E\otimes \mathds{1}_N+\frac{i}{4k}L\otimes \sum_{c=1}^M W_c W_c^\dag,
\end{equation}
with $\sigma_E$ being the shifted $\sigma$ matrix,
\begin{equation}
\label{sgmE}
\sigma_E=\sigma -\frac{E}{4\pi k}\mathds{1}_4.
\end{equation}
As shown in \ref{AppSintegral}, the integral over the supervector $\Psi$ can be done using Eqs. (\ref{gintc}) and (\ref{gintac}), and yields
\begin{equation}
\label{Rksigma}
R_s(k)= \int d[\sigma] \exp(-r\str\sigma^2)
\exp\left(-\frac{i}{4}\mathbf{U}_s^\dag \mathbf{L}^{-1/2}\boldsymbol{\Sigma}^{-1}\mathbf{L}^{-1/2}\mathbf{W}\right)\, \sdet^{-1}\boldsymbol{\Sigma},
\end{equation}
where $\mathbf{L}=L\otimes \mathds{1}_N$ and $\sdet$ represents the superdeterminant~\cite{Efetov1997,Guhr2010}. 
The supersymmetric representation given in Eq.~\eqref{Rksigma} constitutes one of the key results of our paper.
 We have accomplished the difficult task of ensemble averaging and mapped the problem to a matrix integral in superspace. The parameter $N$ which occurred 
 earlier implicitly also, as the dimension of matrix $H$ and supervectors, is now completely an explicit parameter in the integrand. To begin with we had $N^2$ 
 independent integration variables (for $\beta=2$). We now have overall 16 independent integration variables. Thus we have achieved a considerable reduction
  in the number of {\it degrees of freedom} of our problem. This is one of the powerful aspects of the supersymmetry method. 

We now need to analyze the different terms appearing within the integrand in Eq.~\eqref{Rksigma}. As shown in~\ref{AppET}, we find that
\begin{equation}
\label{Lnsdet}
\ln \sdet ^{-1}\bSg=-\str \ln \boldsymbol{\Sigma}=-N \str\ln \sigma_E-\sum_{c=1}^M \str\ln\left(\mathds{1}_4+\frac{i\gamma_c}{4\pi k} 
\sigma_E^{-1}L \right),
\end{equation}
and 
\begin{eqnarray}
\label{Sgminv}
\boldsymbol{\Sigma}^{-1}=\sigma_E^{-1}\otimes\mathds{1}_N-\sigma_E^{-1}\otimes\sum_{c=1}^M \frac{\pi}{\gamma_c}W_c W_c^\dag
+\sum_{c=1}^M \rho^{(c)}\otimes \frac{\pi}{\gamma_c}W_c W_c^\dag,
\end{eqnarray}
where 
\begin{equation}
\label{rho}
\rho^{(c)}=\left(\sigma_E+ \frac{i \gamma_c}{4\pi k} L\right)^{-1}.
\end{equation}
Moreover, the orthogonality relation, Eq.~\eqref{Wortho}, for the vectors $W_c$ enables us to find that (see \ref{AppVLSLW}),
\begin{eqnarray}
\label{VLSLW}
\mathbf{U}_s^\dag \mathbf{L}^{-1/2}\boldsymbol{\Sigma}^{-1}\mathbf{L}^{-1/2}\mathbf{W} 
=\frac{1}{\pi}\left[\gamma_a (-i)^{s+1}\rho_{21}^{(a)}+\gamma_b\, i^{s+1}\rho_{12}^{(b)}\right].
\end{eqnarray}
Thus we see that in this case the exponential factor containing the coupling vectors in Eq.~\eqref{Rksigma} depends only on two matrix elements
 $\rho_{12}^{(c)}$ and $\rho_{21}^{(c)}$ out of the sixteen matrix elements of $\rho^{(c)}$ given in Eq.~\eqref{rho}. 

\subsection{Large-$N$ limit and nonlinear $\sigma$ model}

Our interest is in the limiting case of many resonances $N\gg1$ coupled with few open channels $M \ll N$. Thus we analyze the characteristic function in a 
large-$N$ limit. From the above results we conclude that $R_s(k)$ has the form
\begin{equation}
\label{RkLdL}
R_s(k)=\int d[\sigma] \exp(-N \mathcal{L}-\delta\mathcal{L}),
\end{equation}
where the free energy $\mathcal{L}$ and the perturbation $\delta\mathcal{L}$ around it are given respectively by
\begin{equation}
 \mathcal{L}=r\str \sigma^2+\str \ln \sigma_E
\end{equation}
and
\begin{eqnarray}
\delta\mathcal{L}=\sum_{c=1}^M \str \ln\left(\mathds{1}_4+\frac{i\gamma_c}{4\pi k} \sigma_E^{-1}L \right)
+\frac{i}{4\pi}\left[\gamma_a (-i)^{s+1}\rho_{21}^{(a)}+\gamma_b\, i^{s+1}\rho_{12}^{(b)}\right]. 
\end{eqnarray}
We observe that the term $N\mathcal{L}$ in the free energy is of order $N$ with respect to the term $\delta\mathcal{L}$. The dominating part, $N\mathcal{L}$,
 is invariant under the conjugation by $\mathcal{T}$ which belongs to the Lie superspace $U(1,1/2)$. The sub-dominant part $\delta\mathcal{L}$ breaks this
 symmetry to $U(1/1)\times U(1/1)$. We now fix $M$ and apply the saddle point approximation to consider the $N\rightarrow \infty$ limit. The saddle point
 equation is obtained by the first variation of the exponent as
\begin{equation}
\label{SPeq}
 \sigma_E^{-1}=-2r\sigma,
\end{equation}
which has the diagonal solution
\begin{equation}
 \sigma_D=\frac{E}{8\pi k}\mathds{1}_4+\frac{i \Delta}{8\pi k}  L,
\end{equation}
where $\Delta=(4v^2-E^2)^{1/2}$. We note that $\Delta/(2 \pi v^2)$ is the Wigner semicircle given in Eq.~\eqref{Wsc}. The full solution to Eq. \eqref{SPeq}, 
which contributes to the integral in Eq.~\eqref{RkLdL} in the $N\rightarrow\infty$ limit, is a continuous manifold of the saddle point solutions described by 
\begin{equation}
\label{SPmnfld}
\sigma_G=\frac{E}{8\pi k}\mathds{1}_4-\frac{\Delta}{8\pi k}Q,  
\end{equation}
with $Q=-i \mathcal{T}^{-1}L\mathcal{T}$. $Q$ belongs to the coset superspace U(1,1/2)/[U(1/1)$\times$U(1/1)] and satisfies the conditions $Q^2=-\mathds{1}_4$, 
$\str Q=0$.

We consider the solution as $\sigma=\sigma_G+\delta\sigma$ where $\delta\sigma$ represents the fluctuations around $\sigma_G$. The parts $\sigma_G$ and 
$\delta\sigma$ may be referred to as the Goldstone and massive modes respectively~\cite{SW1980}. Substitution of this solution in Eq.~\eqref{RkLdL} and expansion
 of the terms up to second order in $\delta\sigma$ leads to a separation of the Goldstone modes $\sigma_G$ and the massive modes $\delta\sigma$. The integrals 
over the massive modes are Gaussian ones, and therefore can readily be done and yield unity. We are therefore left with the expression of $R_s(k)$ depending
 on the Goldstone modes $\sigma_G$ only:
\begin{eqnarray}
\label{Rs}
R_s(k)=\int d\mu(\sigma_G) \exp\left[-\frac{i}{4\pi}\big(\gamma_a (-i)^{s+1}\rho_{21}^{(a)}+\gamma_b\, i^{s+1}\rho_{12}^{(b)}\big)\right]
\prod_{c=1}^M\text{sdet}^{-1}\left(\mathds{1}_4+\frac{i\gamma_c}{4\pi k} \sigma_E^{-1}L\right).
\end{eqnarray}
The $\sigma_E$ as well as the $\rho^{(c)}$ in the above equation should now be interpreted as Eqs.~\eqref{sgmE} and \eqref{rho} with $\sigma$ in all the ingredients 
replaced by $\sigma_G$. Eq.~\eqref{Rs} is the nonlinear-sigma model for our problem and constitutes another key result. We would like to emphasize that to arrive
 at this equation we followed a novel route which is based on the characteristic function. This is different from the earlier approaches where one starts with a 
 generating function with source variables. It is also worth mentioning that the superdeterminant part in this equation is the same as those obtained in the earlier
 works~\cite{FS1997}. The exponential part, however, is new in our result and carries the information specific to the present problem.

We now use the parametrization of $Q$ given in~\cite{FS1997}. It involves the pseudo eigenvalues $\lambda_1\in (1,\infty)$, $\lambda_2 \in (-1, 1)$, angles 
$\phi_1, \phi_2 \in (0, 2\pi)$ and four Grassmann variables $\alpha,\alpha^*,\beta,\beta^*$. 
To make the paper self-contained we present this parametrization in ~\ref{AppParam}. As suggested there we write $Q=\mathcal{U}^{-1}\mathcal{D}\mathcal{U}$, and since $\mathcal{U}$ and $L$ commute, the product over the superdeterminant only depends on $\mathcal{D}$. Then it is straightforward to calculate it as 
\begin{equation}
\label{Fu}
\mathcal{F}_\text{U}(\lambda_1,\lambda_2):=\prod_{c=1}^M\text{sdet}^{-1}\left(\mathds{1}_4+\frac{i\gamma_c}{4\pi k} \sigma_E^{-1}L\right)
=\prod_{c=1}^M \frac{g_c^{+}+\lambda_2}{g_c^{+}+\lambda_1},
\end{equation}
which depends only on the pseudo eigenvalues but no other integration variables. Here we defined
\begin{equation}
\label{gcpm}
g_c^{\pm}=\frac{\gamma_c^2\pm v^2}{\gamma_c \Delta},
\end{equation}
where $g_c^{+}$ is related to the transmission coefficient or the sticking probability $T_c=1-|\overline{ S_{cc}}|^2$ as $g_c^{+}=2/T_c-1$~\cite{VWZ1985,FS1997}. The 
transmission coefficient corresponding to a given channel signifies the portion of the flux which is not reflected back immediately, but penetrates the interaction region
and participates in the formation of the long-living resonances~\cite{LW1991}. We will need $g_c^{-}$ later on in the $\beta=1$ case. $\mathcal{F}_\text{U}$ can be 
referred to as the {\it channel factor} since the number of channels $M$ appears explicitly in this term only.

To evaluate the exponential contribution in Eq. \eqref{Rs} we have to explicitly calculate $\rho^{(c)}$ defined in Eq. \eqref{rho} with $\sigma$ replaced by $\sigma_G$. 
This has been done in~\ref{AppRho}. On using the parametrization of~\ref{AppParam} in this result we obtain,
\begin{eqnarray}
\nonumber
\exp\left[-\frac{i}{4\pi}\big(\gamma_a (-i)^{s+1}\rho_{21}^{(a)}+\gamma_b\, i^{s+1}\rho_{12}^{(b)}\big)\right]
&=&\!\!\!\exp \left[\frac{i^{s}k}{2}\left(e^{i\phi_1}t^1_b \left(1-\frac{\alpha^* \alpha}{2} \right)
 \left(1+\frac{\beta^* \beta}{2} \right) - e^{-i\phi_2} t^2_b \alpha^* \beta \right) \right]\\
&\times&\!\!\!\exp \left[-\frac{(-i)^{s} k}{2}  \left(e^{-i\phi_1}t^1_a \left(1-\frac{\alpha^* \alpha}{2} \right)
 \left(1+\frac{\beta^* \beta}{2} \right) + e^{i\phi_2}t^2_a \alpha \beta^* \right)\right].
\end{eqnarray}
We have introduced here
\begin{equation}
\label{tcj}
 t_c^j=\frac{\sqrt{|\lambda_j^2-1|}}{g_c^{+}+\lambda_j};~~~~j=1,2.
\end{equation}
Thus we have the explicit expressions for all the terms in Eq.~\eqref{Rs} in terms of four commuting and four anticommuting variables parametrizing the supermatrix 
$Q$. As we can see, in contrast to the channel factor, the exponential part depends on all integration variables.

\subsection{Reduction to integrals over the radial coordinates}

In this final step we perform the integral over the Grassmann variables and the angles to obtain the results in terms of radial coordinates only. Performing the 
Grassmann integrals amounts to expanding the exponential in the Grassmann variables and picking out the coefficient of $\alpha \alpha^* \beta \beta^*$. All 
terms with other combinations of Grassmann variables vanish. The nonvanishing term turns out to be
\begin{eqnarray}
\nonumber
-\frac{1}{4\pi^2} 
\left[\frac{k^2}{4} t^2_a t^2_b-\frac{k}{8} \left(i^s e^{i\phi_1}t^1_b- (-i)^s e^{-i\phi_1}t^1_a \right)
-\frac{k^2}{16} \left(i^s e^{i\phi_1}t^1_b - (-i)^s e^{-i\phi_1}t^1_a \right)^2\right] 
\exp\left[\frac{k}{2} \left(i^s e^{i\phi_1}t^1_b
- (-i)^s e^{-i\phi_1}t^1_a\right) \right]\\
=-\frac{1}{16\pi^2} 
\left(k^2 t^2_a t^2_b-k\frac{\partial}{\partial k}-k^2\frac{\partial^2}{\partial k^2}\right)\exp\left[\frac{k}{2} \left(i^s e^{i\phi_1}t^1_b
- (-i)^s e^{-i\phi_1}t^1_a\right) \right].~~~~~~~~~~~~~
\end{eqnarray}
 The factor $1/(4\pi^2)$ comes from the convention followed in the definition of Grassmann integration. In the second line of the above equation
 we have recast the expression in front of the exponential as a differential operator acting on the exponential term. We note that there is no $\phi_2$
 dependence in the integrand, the integral over it therefore just gives a factor of $2\pi$. The integral over $\phi_1$ can be performed using the 
following result \cite{GR2000}:
\begin{equation}
 \int_0^{2\pi} d\phi_1\,\exp\left(C_1e^{i\phi_1}-C_2e^{-i\phi_1}\right)=2\pi J_0\left(2\sqrt{C_1 C_2}\right).
\end{equation}
Here $J_0(u)$ represents the zeroth order Bessel function of the first kind. Application of the differential operator on this result then gives the desired expression 
as a 2-fold integral involving Bessel functions  $J_0, J_1, J_2$, which can be further simplified using their recurrence relations. 

It is quite natural to expect a Rothstein or Efetov-Wegner contribution in this result \cite{Efetov1983,Efetov1997,Rothstein1987,KKG2009,Kieburg2011}. It is a 
consequence of the particular choice of the parametrization of the supermatrix and comes from the term of zeroth order in the Grassmann variables, which is 
the product of a divergent result from the integration over the Bosonic variables and zero from the Grassmann integrations. It is known that there is no contribution
 from the second order Grassmannian term in accordance with the results due to Efetov and Zirnbauer~\cite{Zirnbauer1968,Efetov1983}. In our case the 
 Efetov-Wegner contribution is `1' in Eq. \eqref{RkUE} below. It is essential to produce the correct value $R_s(0)=1$ due to the normalization conditions for
 $\mathcal{P}(H)$ and $P_s(x_s)$; see Eqs. \eqref{Rsk}, \eqref{Ps2}. We obtain for both real ($s=1$) and imaginary ($s=2$) parts identical expressions for 
 the characteristic function,
\begin{equation}
\label{RkUE}
R_s(k)=1-\int_1^\infty d\lambda_1\int_{-1}^1 d\lambda_2 \frac{k^2}{4(\lambda_1-\lambda_2)^2}\,\mathcal{F}_\text{U}(\lambda_1,\lambda_2)
 \left(t_a^1 t_b^1+t_a^2 t_b^2\right)J_0\left(k \sqrt{t_a^1 t_b^1}\right).
\end{equation}
The distribution can be obtained using Eq. \eqref{RkUE} in Eq. \eqref{Ps2} and the following Fourier transform results:
\begin{equation}
 \frac{1}{2\pi}\int_{-\infty}^{\infty} dk\, e^{ikx}=\delta(x),
\end{equation}
\begin{equation}
 \frac{1}{2\pi}\int_{-\infty}^{\infty} dk\, e^{ikx} J_0(\omega k) =\frac{1}{\pi\sqrt{\omega^2-x^2}}\Theta\left(\omega^2-x^2\right),
\end{equation}
Here $\Theta(u)$ is the Heaviside-theta function, assuming the value 0 for $u<0$ and 1 for $u>0$.
We obtain
\begin{equation}
\label{PxUE}
 P_s(x_s)=\frac{\partial^2}{\partial x_s^2} f(x_s),
\end{equation}
where
\begin{equation}
\label{fx}
 f(x)=x\Theta(x)+\int_1^\infty d\lambda_1\int_{-1}^1 d\lambda_2 \frac{1}{4\pi(\lambda_1-\lambda_2)^2}\,\mathcal{F}_\text{U}(\lambda_1,\lambda_2)
\big(t_a^1 t_b^1+t_a^2 t_b^2\big)\big(t_a^1 t_b^1-x^2\big)^{-1/2}\Theta(t_a^1 t_b^1-x^2).
\end{equation}
The delta-function singularity at $x_s=0$ in the expression for $P_s(x_s)$ gets canceled by another delta-function singularity hidden in the $\lambda$ integrals. 
These singularities, however, do not create any problem if the evaluation of $P_s(x_s)$ is carried out using Eq. \eqref{PxUE}.

The expression for the $\mu$th moment can be easily obtained from Eq. \eqref{RkUE} by using the series expansion of the Bessel function $J_0$ and examining the
 coefficients of $k^\mu$. We have for $\mu=2n$,
\begin{equation}
\label{evenmom}
 \overline{x_s^{2n}}=\delta_{n,0}+\frac{\Gamma(2n+1)}{2^{2n}\Gamma^2(n)}\int_1^\infty d\lambda_1\int_{-1}^1 d\lambda_2
 \frac{1}{(\lambda_1-\lambda_2)^2}\,\mathcal{F}_\text{U}(\lambda_1,\lambda_2)\left(t_a^1 t_b^1+t_a^2 t_b^2\right)\left|t_a^1 t_b^1\right|^{n-1},
\end{equation}
and for $\mu=2n+1$,
\begin{equation}
\overline{x^{2n+1} }=0,
\end{equation}
where $n=0,1,2,...\,$.

We found above that the distributions of real and imaginary parts are equal in this case. It is therefore clear that the phase $\varphi$ will have a uniform distribution, i.e.,
\begin{equation}
\label{Pphi}
P_\varphi(\varphi)=\frac{1}{2\pi}, 
\end{equation}
 and that the joint density of real and imaginary parts $P_{x_1,x_2}(x_1,x_2)$ will be a function of $r=\sqrt{x_1^2+x_2^2}$ only. The last observation can be used to
 calculate the distribution of modulus also. As shown in the~\ref{AppMod} we obtain for $0<r\leq1$,
\begin{equation}
\label{Pr}
P_r(r)=\frac{1}{r}\frac{\partial}{\partial r}\left(r\frac{\partial}{\partial r}\right) f_r(r),
\end{equation}
where
\begin{equation}
f_r(r)=\frac{1}{2}\frac{(g_a^{+}+\lambda_1)^2 (g_b^{+}+\lambda_1)^2}{(g_a^{+}+g_b^{+})\lambda_1^2+2(g_a^{+} g_b^{+}+1)\lambda_1+ g_a^{+}+g_b^{+}}
\int_{-1}^1 d\lambda_2 \frac{1}{(\lambda_1-\lambda_2)^2}\,\mathcal{F}_U(\lambda_1,\lambda_2)\left[t_a^1 t_b^1+t_a^2 t_b^2\right],
\end{equation}
with $\lambda_1$ assuming the value,
\begin{equation}
\lambda_1=\frac{(g_a^{+}+g_b^{+})r^2+\sqrt{r^2[r^2(g_a^{+}-g_b^{+})^2+4g_a^{+} g_b^{+}-4]+4}}{2(1-r^2)}.
\end{equation}
Note that $P_r(r)$ is normalized as
\begin{equation}
\int_0^1 dr\, r P_r(r)=1.
\end{equation}
As we can see in this case the characteristic function and the distributions have dependence on the parameters of the problem via $g_c^{+}$ (or equivalently the 
transmission coefficients) only. This is similar to the earlier results for $S$-matrix element correlation functions, and distributions of diagonal 
elements~\cite{VWZ1985,FSS2005}. We also note that the cross sections are given by the squared-moduli of the $S$-matrix elements. Consequently we have
access to their distributions also. This is of particular relevance for the experiments where only cross sections are accessible.\\

\begin{figure}[ht]
\centering
\includegraphics[width=0.95\textwidth]{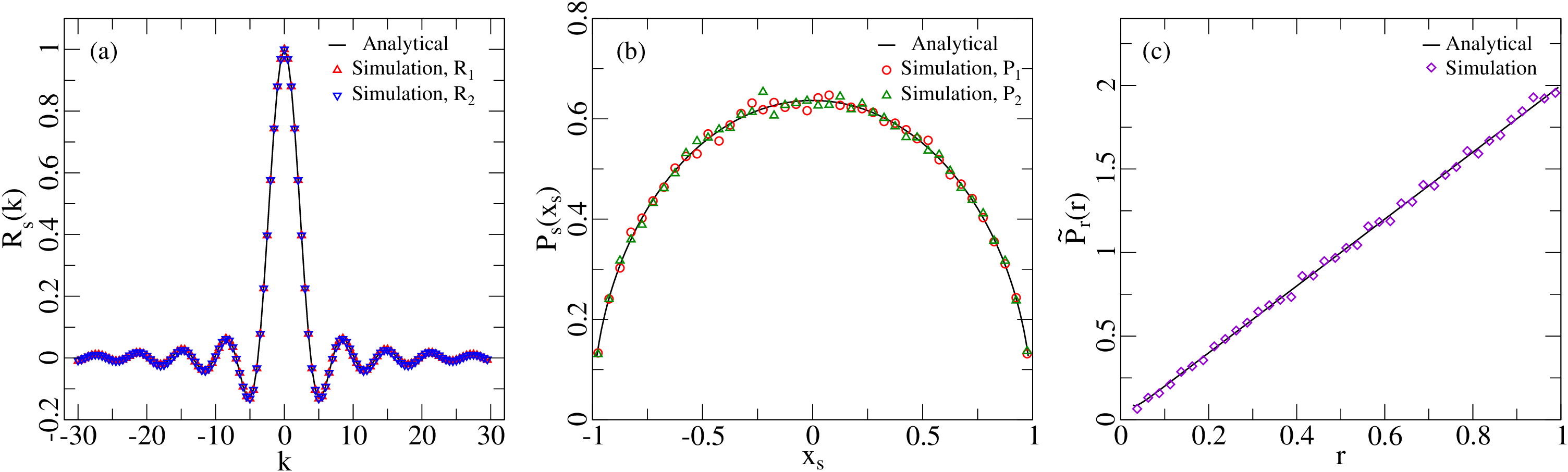}
\caption{Results for $\beta=2$: (a) Characteristic functions $R_s(k)$, (b) Distributions $P_s(x_s)$, and (c) Modulus $\widetilde{P}_r(r)=rP_r(r)$. Values of the
 parameters considered are $M=2, E=0, v=1, \gamma_1=1, \gamma_2=1, a=1, b=2$. }
 \label{Fig1}
\end{figure}
\begin{figure}
\centering
\includegraphics[width=0.95\textwidth]{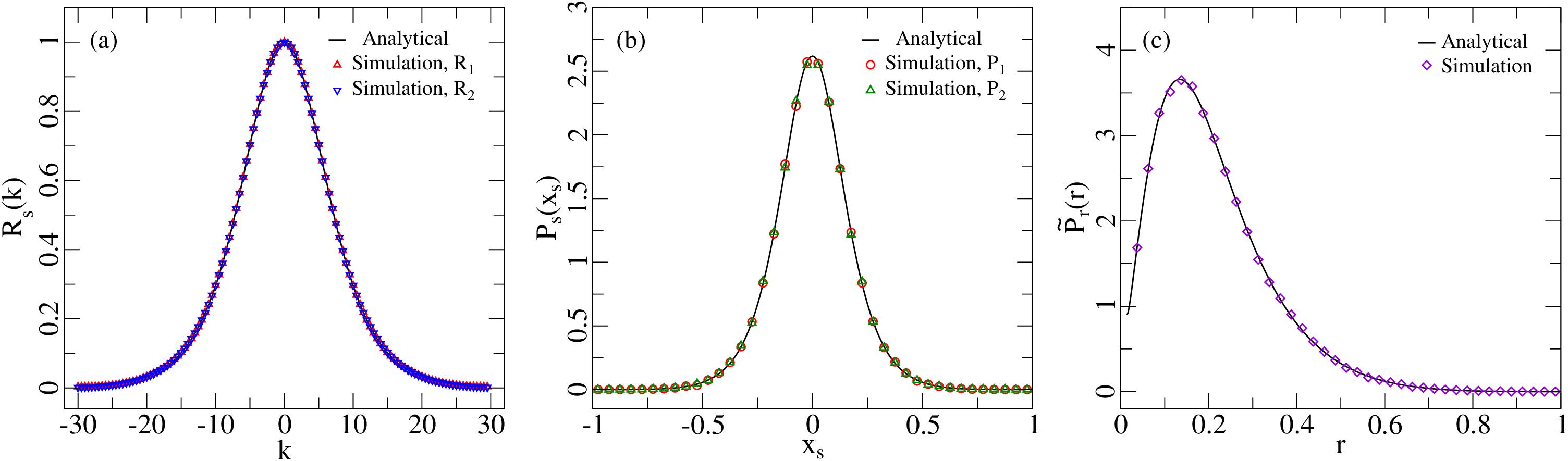}
\caption{Results for $\beta=2$: (a) Characteristic functions $R_s(k)$, (b) Distributions $P_s(x_s)$, and (c) Modulus $\widetilde{P}_r(r)=rP_r(r)$. Values of the
 parameters considered are $M=5, E=1.2, v=1, \gamma_1=0.08, \gamma_2=0.11, \gamma_3=0.27, \gamma_4=0.59, \gamma_5=0.72, a=2, b=3$. }
\label{Fig2}
\end{figure}
\begin{figure}[ht]
\label{Approx}
\centering
\includegraphics[width=0.95\textwidth]{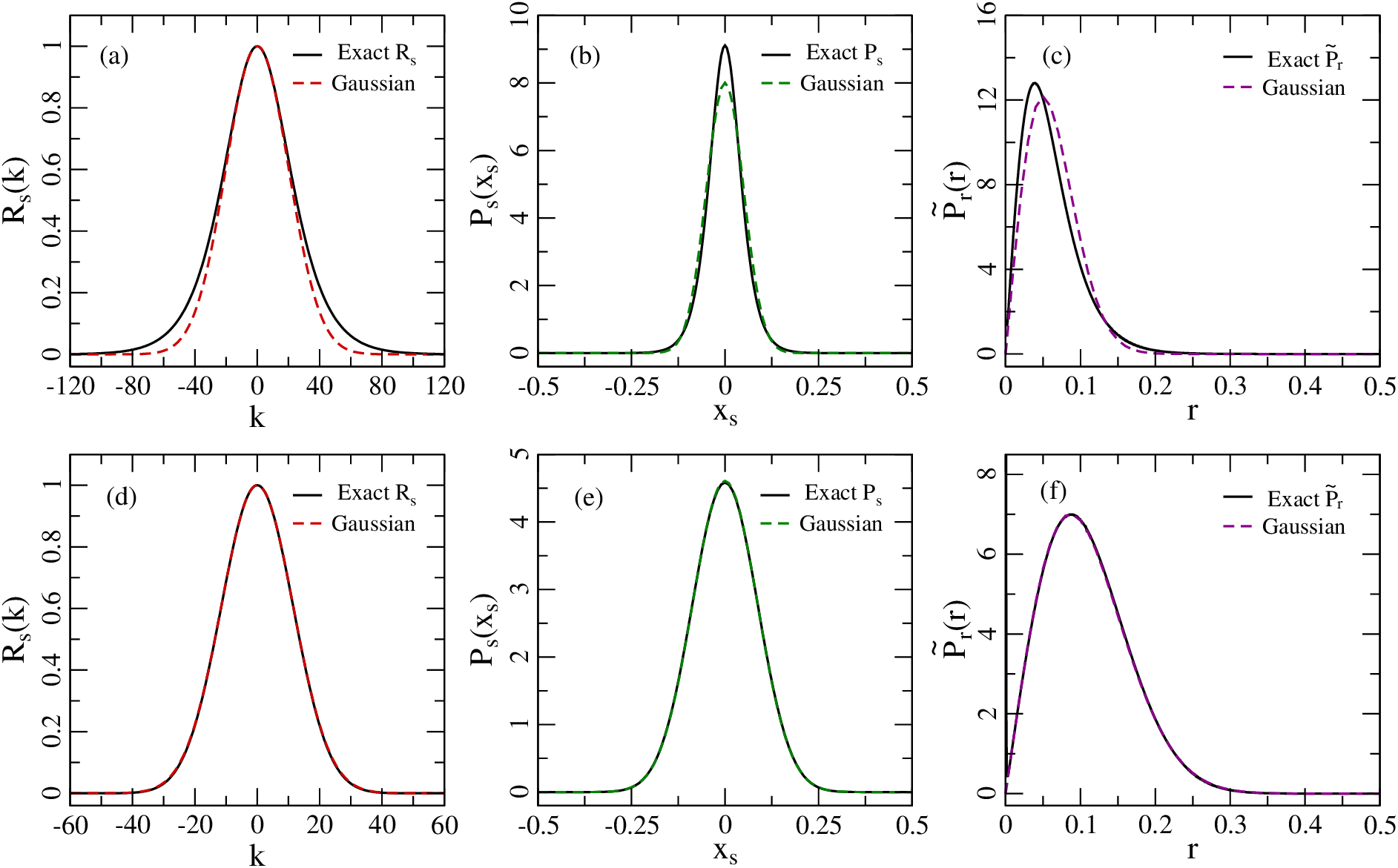}
\caption{Test of the approximate results in Eqs.~\eqref{Psapprox}-\eqref{Rsapprox} for the element $S_{12}$ in the $\beta=2$ case. Figs. (a), (b), (c) show the 
characteristic function, distribution of real (or imaginary) part, distribution of modulus for $\Gamma/\Delta_m\approx0.716$ corresponding to weakly overlapping
 resonances. Figs. (d), (e), (f) depict the same quantities for $\Gamma/\Delta_m\approx8.594$ which is closer to the Ericson regime of strongly overlapping 
 resonances. The solid lines are the exact results while the dashed lines are the approximate ones. The Gaussian approximations work quite well in the regime
  of strongly overlapping resonances.}
\label{Fig3}
\end{figure}
All the above analytical results can be easily implemented in Mathematica~\cite{Mathematica}. The corresponding Mathematica codes can be found as the supplemental material available with Ref.~\cite{NKSG2014}. To test these analytical results we also performed numerical simulations. These simulations were performed with an ensemble of 
 50000 random matrices $H$ of dimensions $250\times250$ from the GUE. For each random matrix we obtain the matrix element $S_{ab}$. For the distributions
 we plot the histogram of such $S$-matrix elements (real part, imaginary part or the modulus) obtained from the ensemble of $H$-matrices. For the characteristic
 functions, instead of obtaining them from the distributions via an inverse Fourier transform, guided by Eq.~\eqref{Rsk}, we use $R_s(k)=(1/n)\sum_{n}\exp(-i k \wp_s(S_{ab}))$, where $n$ represents the number of matrices considered in the ensemble. In Figs.~\ref{Fig1} and \ref{Fig2} we show the plots for (a) the characteristic function, (b) the 
 distributions for the real and imaginary parts and (c) the distribution of the modulus of the scattering matrix element $S_{ab}$. The parameters used for the plots are indicated in the captions. The choice of parameters for Fig.~\ref{Fig1} is such that the transmission coefficient is unity for all the channels (perfect coupling). In this case the $S$ matrix belongs to the Haar measure on the unitary group $U(M)$. In other words it is a member of Dyson's circular unitary ensemble (CUE). In Fig.~\ref{Fig2} we choose values of parameters which 
 corresponds to an ensemble far from the CUE. As we can see the analytical predictions and the numerical simulation results are in excellent agreement in all cases.
 
 It is known that in the case of strongly overlapping resonances (the Ericson regime) the distributions of real and imaginary parts can be well approximated by a 
 Gaussian distribution~\cite{MRW2010},
 \begin{equation}
 \label{Psapprox}
 P_s(x_s)\approx \left(2\pi \mathcal{V}^2\right)^{-1/2}\exp\left(-\frac{x_s^2}{2\mathcal{V}^2}\right),
 \end{equation}
 where the variance of the distribution $\mathcal{V}^2$, which is same as the second moment $\overline{x_s^2}$ (the mean being zero), is determined using 
 Eq.~\eqref{evenmom}. In the same limit the the moduli become Rayleigh distributed, viz.
 \begin{equation}
 \widetilde{P}_r(r)=rP_r(r)\approx \frac{r}{\mathcal{V}^2}\exp\left(-\frac{r^2}{2\mathcal{V}^2}\right).
 \end{equation}
 The corresponding characteristic function is given under this approximation by
 \begin{equation}
 \label{Rsapprox}
 R_s(k)=\exp\left(-\frac{k^2}{2\mathcal{V}^2}\right).
 \end{equation}
 We test these approximations in Fig.~\ref{Fig3}. Figs.~\ref{Fig3} (a), (b), (c) show the characteristic function, distribution of the real (or imaginary) part and distribution
 of the modulus for $M=30$ channels which possess identical values for the transmission coefficient, $T_c=0.15\,(g_c^{+}=12.333)$. Similarly Figs.~\ref{Fig3}
 (e), (f), (g) show these quantities for $M=60$ channels, each having the transmission coefficient $T_c=0.9\, (g_c^{+}=1.222)$. The solid lines are exact results 
 while the dashed lines represent the approximations as in the above equations. In figures (b), (e) and (c), (f), for clarity, the plots have been shown respectively
 for $x_s\in[-0.5,0.5]$ and $r\in[0,0.5]$, instead of [-1,1] and [0,1]. Note that although we focus on the element $S_{12}$, all off-diagonal elements will exhibit 
 the same statistics because of identical choice of $T_c$ for all channels. Using the Weisskopf estimate~\cite{BW1952,DRW2011},
 \begin{equation}
 \label{Weisskopf}
 \frac{\Gamma}{\Delta_m}=\frac{1}{2\pi}\sum_{c=1}^M T_c,
 \end{equation}
 we find that the ratio of the average resonance width ($\Gamma$) and average resonance spacing ($\Delta_m$) is $\Gamma/\Delta_m\approx0.716$ for the former 
 choice and $\Gamma/\Delta_m\approx8.594$ for the latter. The first case corresponds to that of weakly overlapping resonances, while the second one is closer to
 the Ericson regime of strongly overlapping resonances. As expected, we can see significant deviations from the approximate results in Figs.~\ref{Fig3} (a), (b), (c), while 
 in Figs.~\ref{Fig3} (d), (e), (f) the approximations work quite well. 
 
It is worth mentioning that Eq. \eqref{Pr} also provides the exact result for the Landauer conductance of a chaotic quantum dot with two non-ideal leads, each 
supporting a single mode. To see this we recall that the dimensionless Landauer conductance for a chaotic quantum dot supporting $M_1, M_2$, ($M_1+M_2=M$), 
modes in the two leads is given by \cite{Beenakker1997,MK2004,SS1988},
\begin{equation}
\mathcal{G}=\sum_{m=1}^{M_1}\sum_{n=M_1+1}^M |S_{mn}|^2.
\end{equation}
Thus for $M_1=M_2=1$ we have $\mathcal{G}=|S_{12}|^2$, and consequently the distribution of the conductance is the same as the distribution of the modulus-squared of 
the $S$-matrix element $S_{12}$ and can be easily obtained from Eq.~\eqref{Pr}.

\section{Orthogonally Invariant Hamiltonians ($\beta=1$)}
\label{secOE}

We now consider the scenario when $H$ in Eq. \eqref{G} belongs to the GOE, i.e., when it is applicable to systems which are time-reversal as well as rotationally 
invariant. Similar to the unitary case, we again follow three steps to obtain the final results.

 \subsection{Mapping to a matrix integral in superspace}

In the present case $H$ is real symmetric, therefore we have to modify the derivation done for the GUE where $H$ is complex-Hermitian. The reason behind
 this is that if $D$ in Eq. \eqref{ens_av} is left Hermitian and $H$ is taken to be real symmetric, only the real part of $D$ will be affected by the Fourier transform. 
 To obtain $D$ appropriate to the new $H$ we return back to Eq. \eqref{RkA} and carry out the following transformations in $z$ and $\zeta$:
\begin{equation}
z\rightarrow \Xi^{+} z,~~ z^\dag\rightarrow z^\dag, 
~~\zeta\rightarrow 2\,\Xi^{-} \zeta,~~\zeta^\dag\rightarrow\zeta^\dag,
\end{equation}
where $\Xi^{\pm}$ is as defined in Eq.~\eqref{trans}. The Jacobian factor as a result of these transformations is $(-1)^N 2^{-2N}$. Afterwards we decompose the 
$z$ into its real ($x$) and imaginary ($y$) parts to construct a vector double the original size. $x$ and $y$ should not be confused with the real ($x_1$) and imaginary
 ($x_2$) parts of the scattering matrix element. This change from complex to real vectors yields a Jacobian factor of $2^{2N}$ which cancels the same factor generated
  above. We also symmetrize the vector $\zeta$ using $\zeta_a^*, \zeta_b^*$ along with $\zeta_a, \zeta_b$, thus doubling its size as well. Hence we rewrite everything, 
 instead of 4 and $4N$ dimensional objects, in terms of 8 and $8N$ dimensional objects. Moreover, we consider the coupling vectors $W_c$ to be real in this case. 
 We have
\begin{equation}
 R_s(k)=(-1)^N \int d[\Psi] \exp\left(i\Psi^\dag \mathbf{V}_s\right)\int d[H] 
\mathcal{P}(H)\exp\left(\frac{i}{4\pi k} \Psi^\dag \boldsymbol{\mathcal{A}}^{-1} \Psi\right),
\end{equation}
As clear from the above discussion, we now have
\begin{equation}
\Psi=\begin{bmatrix}
      x_a \\ y_a \\ x_b \\ y_b \\ \zeta_a \\ \zeta_a^* \\ \zeta_b \\ \zeta_b^*
\end{bmatrix},
~~\Psi^\dag=\begin{bmatrix} x_a^T & y_a^T & x_b^T & y_b^T & \zeta_a^\dag & -\zeta_a^T & \zeta_b^\dag & -\zeta_b^T\end{bmatrix};
~~~\mathbf{V}_s=\frac{1}{2}\begin{bmatrix} W_a-i^s W_b\\-i (W_a+i^s W_b)\\(-i)^sW_a+W_b\\i((-i)^sW_a-W_b)\\0 \\ 0 \\ 0 \\0  \end{bmatrix},
\end{equation}
and $\boldsymbol{\mathcal{A}}^{-1}=\text{diag}(-(G^{-1})^\dag,G^{-1},-(G^{-1})^\dag,-G^{-1})\otimes\mathds{1}_2$. With these modifications Eq. \eqref{ens_av}
 holds  with
\begin{equation}
D=x_a x_a^T+y_a y_a^T-x_b x_b^T-y_b y_b^T-\zeta_a \zeta_a^\dag+\zeta_a^* \zeta_a^T-\zeta_b \zeta_b^\dag+\zeta_b^* \zeta_b^T.
\end{equation}
and
\begin{equation}
r=\frac{4\pi^2k^2 N}{v^2}.
\end{equation}
We note that $D$ is now real symmetric. Eqs. (\ref{Bmn}) and (\ref{trstr}) are also applicable, but now with  $K=\text{diag}(1,1,-1,-1,1,1,1,1)$ and $m,n=1,...,8$. 
Moreover, we have $\Psi_1\equiv x_a,\Psi_2\equiv y_a,\Psi_3\equiv x_b,\Psi_4\equiv y_b,\Psi_5\equiv\zeta_a$, $\Psi_6\equiv\zeta_a^*$,
$\Psi_7\equiv\zeta_b$, and $\Psi_8\equiv\zeta_b^*$. Thus, analogous to Eq. \eqref{RkLBL_UE}, we obtain
\begin{equation}
R_s(k)=(-1)^N \int d[\Psi] \exp\left[-\frac{1}{4r}\str (K^{1/2} \mathcal{B} K^{1/2})^2\right]
 \exp\left[\frac{i}{4\pi k} \Psi^\dag \boldsymbol{\mathcal{A}}_0^{-1} \Psi +i\Psi^\dag \mathbf{V}_s\right],
\end{equation}
where $\boldsymbol{\mathcal{A}}_0^{-1}$ is given by Eq. \eqref{A0_UE} with $L=\diag(1,1,-1,-1,1,1,-1,-1)$, $\1_4\rightarrow \1_8$, and $K$ as defined above. 
With the above considerations we accomplish the GOE averaging. Similar to the unitary case we now use the Hubbard-Stratonovich identity, Eq. \eqref{HSid}, 
with an $8\times 8$ dimensional supermatrix $\sigma$~\cite{VWZ1985}. The integral over the supervector $\Psi$ in this case is a lot more involved than 
that for $\beta=2$. The main steps are outlined in~\ref{AppSintegral}. We arrive at
\begin{equation}
\label{RksigmaOE}
R_s(k)= \int d[\sigma] \exp\left(-r\str\sigma^2\right)
\exp\left({-\frac{i}{4}\mathbf{V}_s^T \mathbf{L}^{-1/2}\bSg^{-1}\mathbf{L}^{-1/2}\mathbf{V}_s}\right)
\text{sdet}^{-1/2}\bSg.
\end{equation}
The quantities in the integrand can be read out from Eqs.~\eqref{Lnsdet}-\eqref{rho} taking into consideration the modifications mentioned above. 

The orthogonality relation, Eq.~\eqref{Wortho},  of the vectors $W_c$  enables us to find that (similar to the $\beta=2$ calculation in~\ref{AppVLSLW})
\begin{eqnarray}
\label{exp1}
\nonumber
\mathbf{V}_s^T \mathbf{L}^{-1/2}\bSg^{-1}\mathbf{L}^{-1/2}\mathbf{V}_s 
= \frac{\gamma_a}{4\pi} \Big[      \rho_{11}^{(a)}          -i \rho_{12}^{(a)}
                          +(-i)^{s+1} \rho_{13}^{(a)}     +(-i)^s \rho_{14}^{(a)}
                                   -i \rho_{21}^{(a)}           - \rho_{22}^{(a)}
                              -(-i)^s \rho_{23}^{(a)} +(-i)^{s+1} \rho_{24}^{(a)}  \\
\nonumber
                          +(-i)^{s+1} \rho_{31}^{(a)} -(-i)^s \rho_{32}^{(a)}
                               -(-1)^s \rho_{33}^{(a)}   -i(-1)^s \rho_{34}^{(a)}    
                             +(-i)^s \rho_{41}^{(a)}  +(-i)^{s+1} \rho_{42}^{(a)}
                             -i(-1)^s \rho_{43}^{(a)}     +(-1)^s \rho_{44}^{(a)} \Big]\\
\nonumber
  +\frac{\gamma_b}{4\pi} \Big[  (-1)^s \rho_{11}^{(b)}     +i(-1)^s \rho_{12}^{(b)}
                                 +i^{s+1} \rho_{13}^{(b)}         +i^s \rho_{14}^{(b)}
                                 +i(-1)^s \rho_{21}^{(b)}
                                 -(-1)^s \rho_{22}^{(b)}         -i^s \rho_{23}^{(b)} 
                                 +i^{s+1} \rho_{24}^{(b)}  \\
				   +i^{s+1} \rho_{31}^{(b)}
                                     -i^s \rho_{32}^{(b)}            - \rho_{33}^{(b)}  
                                      +i \rho_{34}^{(b)}         +i^s \rho_{41}^{(b)}
                                 +i^{s+1} \rho_{42}^{(b)}           +i \rho_{43}^{(b)}
                                        + \rho_{44}^{(b)} \Big].~~
\end{eqnarray}
~\\
The complexity of the calculation in the $\beta=1$ case over that in $\beta=2$ can be realized from the above expression. Earlier we had just 2 elements of $\rho^{(c)}$
in the corresponding term (cf. Eq. \eqref{VLSLW}), now we have 32 (16+16) elements coming from the Boson-Boson blocks of $\rho^{(a)}$ and $\rho^{(b)}$. 

\subsection{Large-$N$ limit and nonlinear $\sigma$ model}

Similar to the $\beta=2$ case, for the $N\rightarrow\infty$ limit, we implement the saddle-point approximation as described by Eqs. \eqref{SPeq}-\eqref{SPmnfld},
the difference being that we now deal with $8$-dimensional supermatrices. The integration over the massive modes again are Gaussian ones and give just a factor
of unity. The matrix $Q$ in this case belongs to the coset superspace UOSP(2,2/4)/[UOSP(2,2)$\times$UOSP(2/2)].  

Next we express the result in terms of the parametrization given in the~\ref{AppParam}. This involves three pseudo eigenvalues $\lambda_0\in(-1,1),
\lambda_1,\lambda_2\in(1,\infty)$, two O(2) angles $\phi_1,\phi_2\in(0,2\pi)$, three SU(2) variables $m,r,s\in(-\infty,\infty)$, and eight Grassmann variables. 
It turns out that the Boson-Boson block of $\rho$ is symmetric, i.e., $\rho_{ij}=\rho_{ji},\ i,j=1...4$. This reduces the number of independent elements of
 $\rho^{(c)}$ entering the exponent, Eq.~\eqref{exp1}, to 20.

The evaluation of the product over the superdeterminant part does not pose much difficulty and turns out to be dependent only on the pseudo eigenvalues. 
Using similar arguments as in the unitary case, we obtain
\begin{equation}
\label{Fo}
\mathcal{F}_\text{O}(\lambda_0,\lambda_1,\lambda_2):=\prod_{c=1}^M\text{sdet}^{-1/2}\left(\mathds{1}_8+\frac{i\gamma_c}{4\pi k} \sigma_E^{-1}L\right)
=\prod_{c=1}^M \frac{g_c^{+} + \lambda_0}
{(g_c^{+} + \lambda_1)^{1/2}(g_c^{+} + \lambda_2)^{1/2}},
\end{equation}
where $g_c^{+}$ is as defined in Eq.~\eqref{gcpm}. The evaluation of the exponential term, however, requires elaborate calculations as described below. 

\subsection{Reduction to integrals over the radial and angular coordinates}

The integration over the Grassmann variables is extremely challenging in this case. The complexity can be understood from the fact that in this case we have
 8 Grassmann variables, which can give rise to a total of 127 terms which have even number of Grassmannian variables in them, i.e., terms with two, four, six 
 or eight Grassmann variables, in contrast to just 7 possible terms in the $\beta=2$ case. Apart from these we have one term with no Grassmannian (zeroth order term). 
 To perform the integral we have to look for all the combinations which give rise to terms with all 8 Grassmannian variables when multiplied together. The above 
 described 128 terms lead to a total of 379 possible combinations. It turns out that in our actual calculation we have 100 terms with even number of Grassmann 
 variables in them. These terms lead to 226 combinations which consist of all eight Grassmann variables. It is evident that these calculations are extremely
 lengthy and cumbersome to be performed manually. We therefore used Mathematica~\cite{Mathematica} to accomplish this task. The Grassmann algebra was
  performed using the {\it grassmann}~\cite{MH} and {\it grassmannOps}~\cite{JM} packages for Mathematica. 

After integrating out these Grassmann variables, we are left with an integral over the 8 commuting variables. The resultant expression being extremely lengthy
we refrain from presenting it here. Remarkably enough, this expression when carefully simplified, does not contain the SU(2) variables $m,r,s$ at all; they appear
 only in the Jacobian (see~\ref{AppParam}). 
Thus the integral over them can be trivially performed, leading to just a factor of $\pi^2$, i.e.,
\begin{equation}
\int_{-\infty}^\infty dm \int_{-\infty}^\infty dr \int_{-\infty}^\infty ds\, \frac{1}{(1+m^2+r^2+s^2)^2}=\pi^2.
\end{equation}
We are therefore left with two O(2) variables $\phi_1,\phi_2$ and three pseudo-eigenvalues $\lambda_0, \lambda_1, \lambda_2$. In general, for $\beta=1$ 
supersymmetry calculations one is able to obtain the final result in terms of a three-fold integral involving the pseudo eigenvalues. However, in this case the 
integrand has a very complicated dependence on the O(2) variables which has the form
\begin{equation}
\label{cfOE}
F(\lambda_0,\lambda_1,\lambda_2,\phi_1,\phi_2)\exp\left[A_1 e^{i(\phi_1-\phi_2)}+A_2 e^{-i(\phi_1-\phi_2)}+B_1 e^{i2\phi_1}+B_2 e^{-i2\phi_1}
+C_1 e^{i2\phi_2}+C_2 e^{-i2\phi_2}\right],
\end{equation}
where $F$ is a complicated multi-term expression involving the $\lambda$'s and the $\phi$'s. In particular the $\phi_i$ occur in the form $e^{i(n_1\phi_1+n_2\phi_2)}$, 
where $n_1,n_2$ are integers. Because of these specific $\phi$-dependences we can rewrite $F$ as a differential operator acting on the $\phi_1, \phi_2$ integral 
having integrand as the exponential factor in the above equation. By going to center and difference variables $\varphi=(\phi_1+\phi_2)/2$ and $\psi=\phi_1-\phi_2$
 we are then able to perform one more integral (over $\varphi$ ), leaving the remaining $\psi$-integral as
\begin{equation}
\label{BessInt}
2\pi\int_0^{2\pi}  d\psi\, I_0\left(2\sqrt{(A_1+B_1 e^{i2\psi}+C_2 e^{-i2\psi})(A_2+B_2 e^{-i2\psi}+C_1 e^{i2\psi})}\right),
\end{equation}
where $I_0(u)$ represents the modified Bessel function of the first kind and is related to the Bessel function $J_0(u)$ as $I_0(i u)=J_0(u)$. Unfortunately, there
does not seem to be any closed form result for this integral. Thus the final expression of $R_s(k)$ is given as a four-fold integral. 

We obtain $R_s(k)$ as, 
\begin{equation}
\label{RkOE}
R_s(k)=1+\frac{1}{8\pi}\int_{-1}^1 d\lambda_0 \int_1^\infty d\lambda_1\int_1^\infty d\lambda_2 \int_0^{2\pi}d\psi 
~\mathcal{J}(\lambda_0,\lambda_1,\lambda_2)\mathcal{F}_O\left(\lambda_0,\lambda_1,\lambda_2\right)\left(\kappa_1 k+\kappa_2 k^2+\kappa_3 k^3
+\kappa_4 k^4\right).
\end{equation}
The `1' in the above equation is again a consequence of the Efetov-Wegner correction. The Jacobian factor is given by
\begin{equation}
\mathcal{J}(\lambda_0,\lambda_1,\lambda_2)=\frac{(1-\lambda_0^2)|\lambda_1-\lambda_2|}
{2(\lambda_1^2-1)^{1/2}(\lambda_2^2-1)^{1/2}(\lambda_1-\lambda_0)^2(\lambda_2-\lambda_0)^2},
\end{equation}
while the channel factor $\mathcal{F}_\text{O}$ is defined as in Eq. \eqref{Fo}. The fourth degree polynomial in $k$ with coefficients $\kappa$'s in \eqref{RkOE}
 is a consequence of the application of the above described differential operator to the Bessel function appearing in Eq.~\eqref{BessInt}. To define these $\kappa$'s
 we need the following:
\begin{equation}
p_c^j=\frac{\sqrt{|\lambda_j^2-1|}}{8(g_c^{+} +\lambda_j)},~~~~ j=0,1,2,
\end{equation}
\begin{equation}
p_c^{\pm}=p_c^1\pm p_c^2,
\end{equation}
\begin{equation}
\label{qcp}
q_c^{+}=\frac{(-i)^{s-1}}{8}\left(\frac{E}{\Delta}+i g_c^{-}\right)\left(\frac{1}{g_c^{+}+\lambda_1}+\frac{1}{g_c^{+}+\lambda_2}
-\frac{2}{g_c^{+}+\lambda_0}\right),
\end{equation}
\begin{equation}
\label{qcm}
q_c^{-}=\frac{(-i)^{s-1}}{8}\left(\frac{E}{\Delta}+i g_c^{-}\right)\left(\frac{1}{g_c^{+}+\lambda_1}-\frac{1}{g_c^{+}+\lambda_2}\right).
\end{equation}
Recall that $s=1,2$ correspond respectively to the real and imaginary parts of $S_{ab}$. We also consider the complex conjugate of $q_c^{\pm}$, $r_c^{\pm}=(q_c^{\pm})^*$, 
and the quantities $l=X/Y, m=Y/X, \omega=2\sqrt{X Y}$, where
\begin{equation}
X=2p_a^{+}+q_a^{-}e^{-i2\psi}+r_a^{-}e^{i2\psi},~~~~
Y=2p_b^{+}-q_b^{-}e^{i2\psi}-r_b^{-}e^{-i2\psi}.
\end{equation}
It can be verified that $\omega^2$ is real for all the values of parameters involved and assumes the values from 0 to 1. The $\kappa$'s are given as
\begin{eqnarray}
\nonumber
 \kappa_1=\kappa_{11} J_1(k\omega),
~~~ 
\kappa_2=\kappa_{21} J_0(k \omega)+\kappa_{22} J_2(k \omega),\hspace{1.8cm}\\
 \kappa_3=\kappa_{31} J_1(k\omega)+\kappa_{32} J_3(k\omega),
~~~
 \kappa_4=\kappa_{41} J_0(k\omega)+\kappa_{42} J_2(k\omega)+\kappa_{43} J_4(k\omega),
\end{eqnarray}
~\\
The coefficients with the Bessel functions above are as follows
\begin{equation}
\kappa_{11}=-(9/8)\{p_a^{+}m^{1/2} \}_{+},
\end{equation}
\begin{eqnarray}
\kappa_{21}=-(1/4)(128 p_a^0p_b^0+14p_a^{+}p_b^{+}+32p_a^{-}p_b^{-})
+\{3e^{i2\psi}(p_a^{-}q_b^{+}-p_b^{-}r_a^{+})\}_{-}
+\{e^{-4i\psi}q_a^{-}r_b^{-}\}_{+},
\end{eqnarray}
\begin{equation}
\kappa_{22}=-(1/4)\{(p_a^{+}p_a^{+}-4q_a^{-}r_a^{-})m\}_{+}, 
\end{equation}
\begin{eqnarray}
\nonumber
&&\kappa_{31}=\big\{2\big[(p_a^{+}p_a^{+}+q_a^{-}r_a^{-})m^{1/2}+2(8p_a^0p_b^0+p_a^{+}p_b^{+}+p_a^{-}p_b^{-})l^{1/2}\big](e^{i2\psi}q_b^{-}+e^{-i2\psi}r_b^{-})\big\}_{-}\\
\nonumber
&&+\big\{2\big[(p_a^{+}p_b^{-}+4p_a^{-}p_b^{+})m^{1/2}+p_b^{+}p_b^{-}l^{1/2}\big](e^{-i2\psi}q_a^{+}+e^{i2\psi}r_a^{+})\big\}_{-}
+\big\{\big[16p_a^0(2p_a^0 p_b^{+}-3p_b^0 p_a^{+})\\
\nonumber
&&
-6p_a^{+}(q_a^{+} q_b^{+}+r_a^{+} r_b^{+})+2p_b^{+}(4q_a^{+} r_a^{+}-q_a^{-} r_a^{-})-4p_a^{-}(p_a^{+}p_b^{-}-2p_a^{-}p_b^{+})+3p_a^{+}(q_a^{-}q_b^{-}+r_a^{-}r_b^{-}-p_a^{+}p_b^{+})
\\
\nonumber
&&+(e^{-i4\psi}/2)q_a^{-}(4p_a^{+}r_b^{-}-3p_b^{+}q_a^{-})+(e^{i4\psi}/2)r_a^{-}(4p_a^{+}q_b^{-}-3p_b^{+}r_a^{-})\big]m^{1/2}\big\}_{+}\\
&&+\big\{\big[(e^{-i4\psi}/2)q_a^{-}(2e^{-i2\psi}q_a^{-}r_b^{-}-8e^{i2\psi}r_a^{+}r_b^{+})+(e^{i4\psi}/2)r_a^{-}(2e^{i2\psi}q_b^{-}r_a^{-}-8e^{-i2\psi}q_a^{+}q_b^{+})\big]m^{1/2}\big\}_{-},
\end{eqnarray}
\begin{eqnarray}
 \kappa_{32}=\big\{p_a^{+}\big[(p_a^{+}p_a^{+}+2q_a^{-}r_a^{-})
 +(3/2)(e^{-i4\psi} q_a^{-}q_a^{-}+e^{i4\psi}r_a^{-}r_a^{-})\big]m^{3/2}\big\}_{+}
 +\big\{(2p_a^{+}p_a^{+}+q_a^{-}r_a^{-})(e^{-i2\psi} q_a^{-}+e^{i2\psi}r_a^{-})m^{3/2}\big\}_{-},
\end{eqnarray}
\begin{eqnarray}
\nonumber
 \kappa_{41}\!\!\!&=&\!\!\!32 \big[2 p_a^0 p_a^0(p_b^{-}-e^{i2\psi}q_b^{+})(p_b^{-}-e^{-i2\psi}r_b^{+})
   +2 p_b^0 p_b^0(p_a^{-}+e^{-i2\psi}q_a^{+})(p_a^{-}+e^{i2\psi}r_a^{+})\\
     \nonumber
   &+&\!\!\! p_a^0 p_b^0\big((p_a^{+}+e^{-i2\psi}q_a^{-})(p_b^{+}-e^{-i2\psi}r_b^{-})+(p_a^{+}+e^{i2\psi}r_a^{-})(p_b^{+}-e^{i2\psi}q_b^{-})\big)\big]\\
   \nonumber
   &+&\!\!\! 256 p_a^0 p_a^0p_b^0 p_b^0
   +(p_a^{+}+e^{-i2\psi}q_a^{-})^2(p_b^{+}-e^{-i2\psi}r_b^{-})^2
   +(p_a^{+}+e^{i2\psi}r_a^{-})^2(p_b^{+}-e^{i2\psi}q_b^{-})^2\\
    \nonumber
   &+&\!\!\! 4\big[(p_a^{+}+e^{-i2\psi}q_a^{-})(p_b^{+}-e^{i2\psi}q_b^{-})
-2(p_a^{-}+e^{-i2\psi}q_a^{+})(p_b^{-}-e^{i2\psi}q_b^{+})\big]\\
   &&\times\big[(p_a^{+}+e^{i2\psi}r_a^{-})(p_b^{+}-e^{-i2\psi}r_b^{-})
-2(p_a^{-}+e^{i2\psi}r_a^{+})(p_b^{-}-e^{-i2\psi}r_b^{+})\big],
\end{eqnarray}
\begin{eqnarray}
\nonumber
 \kappa_{42}=-32p_a^0p_b^0\big[(p_a^{+}+e^{-i2\psi}q_a^{-})(p_a^{+}+e^{i2\psi}r_a^{-})m
 +(p_b^{+}-e^{i2\psi}q_b^{-})
(p_b^{+}-e^{-i2\psi}r_b^{-}) l ]\hspace{2cm}\\
 \nonumber
- 2\big[(p_a^{+}+e^{-i2\psi}q_a^{-})(p_b^{+}-e^{i2\psi}q_b^{-})
-2(p_a^{-}+e^{-i2\psi}q_a^{+})(p_b^{-}-e^{i2\psi}q_b^{+})\big]
\big[(p_a^{+}+e^{i2\psi}r_a^{-})^2 m
 +(p_b^{+}-e^{-i2\psi}r_b^{-})^2 l\big]\\
 - 2\big[(p_a^{+}+e^{i2\psi}r_a^{-})(p_b^{+}-e^{-i2\psi}r_b^{-})
-2(p_a^{-}+e^{i2\psi}r_a^{+})(p_b^{-}-e^{-i2\psi}r_b^{+})\big]
\big[(p_a^{+}+e^{-i2\psi}q_a^{-})^2 m
 +(p_b^{+}-e^{i2\psi}q_b^{-})^2 l \big],
\end{eqnarray}
\begin{eqnarray}
 \kappa_{43}=(p_a^{+}+e^{-i2\psi}q_a^{-})^2(p_a^{+}+e^{i2\psi}r_a^{-})^2 m^2
 +(p_b^{+}-e^{i2\psi}q_b^{-})^2(p_b^{+}-e^{-i2\psi}r_b^{-})^2 l^2.
\end{eqnarray}
In the above equations,  an expression $\mathcal{E}$ involving $a,b,l,m,\psi$ enclosed in the bracket $\{\,\}_{\pm}$ represents $\{\mathcal{E}(a,b,l,m,\psi)\}_{\pm}
:=\mathcal{E}(a,b,l,m,\psi)\pm\mathcal{E}(b,a,m,l,-\psi)$. As we can see the distributions for the real and imaginary parts are not the same in this case. They differ 
by the definition of $q_c^{\pm}$ in Eqs. \eqref{qcp}, \eqref{qcm}. This explains the unequal deviations of the real and imaginary parts from a Gaussian behavior which
was observed in~\cite{T1975,RSW1975} but could not be understood.
 
The expression for $R_s(k)$, Eq. \eqref{RkOE}, exhibits several symmetries. The $k\rightarrow -k$ symmetry is easily visible. The symmetry $a\leftrightarrow b$, 
as expected for the $\beta=1$ case, is hidden because of the remaining $\psi$-integral. The $\psi$-integral being periodic, the transformation 
$\psi\rightarrow -\psi\pm \pi/2$ does not alter the value of the integral. The transformed integrand then exhibits $a\leftrightarrow b$ symmetry with the original integrand. 
$E\rightarrow -E$ symmetry is revealed once one performs the transformations $a\leftrightarrow b$ and $E\rightarrow -E$ together, as then one returns back to the 
original expression. Note that for $g_c^{+}=1$ (obtained for $v=1, E=0,\gamma_c=1$ for all $c$), corresponding to Dyson's Circular Orthogonal Ensemble (COE), 
$q_c^{\pm}$ and hence $r_c^{\pm}$ become zero, and also the $\psi$-dependence goes away from the integrand. The expression then simplifies a lot and the 
$\psi$-integral just gives a value of $2\pi$ resulting in a final three-integral expression. Moreover, the result then becomes identical for the real and imaginary parts.

A new feature emerging in the above expression, compared to the $\beta=2$ case, is the explicit dependence on quantities other than $g_c^{+}$ (and thus the
transmission coefficient $T_c$), namely $E/\Delta$ and $g_c^{-}$. Note that these quantities are relevant only for the channels $a$ and $b$ if we are interested
in the statistics of the matrix element $S_{ab}$. The information about the rest of the channels ($c\neq a,b$) enter only via the channel factor which just involves 
$g_c^{+}$. To demonstrate this extra dependence let us consider $g_c^+=g$ for all $c$. In this case, for fixed $\gamma$ and $E$, we have two choices for $v$ 
given by $v^2=\gamma^2(2g^2-1)\pm\gamma g\sqrt{4\gamma^2(g^2-1)-E^2}$, provided $4\gamma^2(g^2-1)-E^2>0$. For these two choices, the parameters
 $E/\Delta$ and $g_c^{-}$ have different values and lead to different characteristic functions, and hence different distributions for the real and imaginary parts. If
 we fix the scale $v^2$ and the energy $E$, we are still left with the $\gamma_c\leftrightarrow v^2/\gamma_c$ duality, i.e., the choices $\gamma_c=\gamma$ or
 $v^2/\gamma$ lead to the same $g_c^{+}$, but $g_c^{-}$ with an opposite sign. However, if this duality is implemented simultaneously for the channels $a,b$ then
 the results remain unchanged. This is a consequence of $E\leftrightarrow -E$, $a\leftrightarrow b$ and $\psi\leftrightarrow -\psi$ symmetries.
 If the $\gamma_c\leftrightarrow v^2/\gamma_c$ change is implemented in only one of the $a, b$ channels, the results change. This extra freedom may seem surprising 
 at first, however a careful examination reveals that while $g_c^{+}$ determines the absolute value of $\overline{S_{cc}}$,  the quantity $\tan^{-1}((E/\Delta)/g_c^{-})$
 determines the phase of $\overline{S_{cc}}$. Also, $(E/\Delta)^2+(g_c^{-})^2=(g_c^{+})^2-1$. Consequently, $g_c^{+}$ and the complex quantity $E/\Delta+i g_c^{-}$
 together contain no more information than that contained in $\overline{S_{cc}}$. Thus, unlike the $\beta=2$ case, where the characteristic functions and distributions 
 of the real and imaginary parts of a given off-diagonal element $S_{ab}$ are completely determined by the absolute value of $\overline{S_{cc}} \,(c=1,...,M)$, for 
 $\beta=1$ we additionally need the information about the phases for $\overline{S_{aa}}$ and $\overline{S_{bb}}$.
 
The distributions can be written, upon taking the Fourier transform of $R_s(k)$, as expressions containing up to fourth-derivative with respect to $x_s$. We have
\begin{equation}
\label{Px}
P_s(x_s)=\delta(x_s)+\frac{\partial f_1}{\partial x_s}+\frac{\partial^2 f_2}{\partial x_s^2}+\frac{\partial^3 f_3}{\partial x_s^3}+\frac{\partial^4 f_4}{\partial x_s^4},
\end{equation}
where
\begin{eqnarray}
\nonumber
&&f_1=\left\langle\kappa_{11} x_s/\omega
\right\rangle,\\
\nonumber
&&f_2=-\left\langle\kappa_{21}+\kappa_{22}\big(1-2x_s^2/\omega^2\big)
\right\rangle,\\
\nonumber
&&f_3=-\left\langle\big[\kappa_{31}+\kappa_{32}\big(3-4x_s^2/\omega^2\big)\big]x_s/\omega\right\rangle,\\
&&f_4=\big\langle\big[\kappa_{41}+\kappa_{42}\big(1-2x_s^2/\omega^2\big)+\kappa_{43}\big(1-8x_s^2/\omega^2+8x_s^4/\omega^4\big)\big]
\big\rangle.
\end{eqnarray}
~\\
Here the angular brackets represent the following:
\begin{equation}
\langle h \rangle=\frac{1}{16\pi^2}\int_{-1}^1 \!d\lambda_0 \int_1^\infty \!d\lambda_1\int_1^\infty \!d\lambda_2 \int_0^{2\pi}\!d\psi \mathcal{J}(\lambda_0,\lambda_1,\lambda_2)
\mathcal{F}_\text{O}(\lambda_0,\lambda_1,\lambda_2)\, 2h \left(\omega^2-x_s^2\right)^{-1/2}\Theta\left(\omega^2-x_s^2\right).
\end{equation}

\begin{figure}[t]
\centering
\includegraphics*[width=0.7\textwidth]{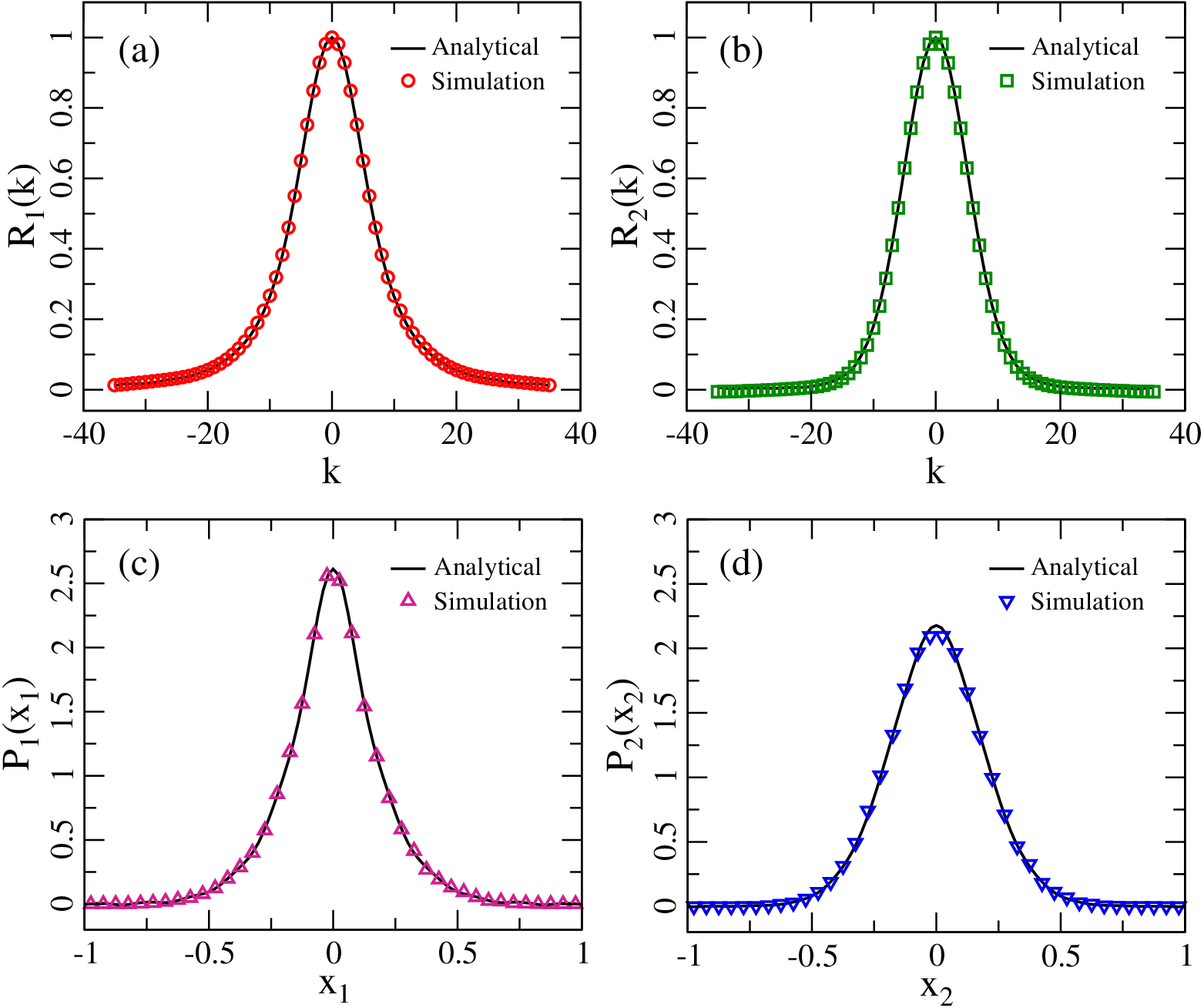}
\caption[]{Results for $\beta=1$: Characteristic function for (a) real part, (b) imaginary part; Distribution for (c) real part, (d) imaginary part. 
Values of the parameters considered are $M = 6, E = 0.75, v = 1, \gamma_1 = 0.2, \gamma_2 = 1.3, \gamma_3 = 0.17, \gamma_4 = 0.5, \gamma_5 = 0.9, 
\gamma_6=1.1, a = 3, b = 4$. Solid lines represent the analytical 
predictions while the symbols are from simulations.}
\label{Fig4}
\end{figure}

The analytical result for the characteristic function, Eq.~\eqref{RkOE}, can be implemented in Mathematica~\cite{Mathematica}. 
We found that the Efetov variables $\theta_j, j=0,1,2$ \cite{Efetov1983} are best suited for the numerical evaluation of $R_s(k)$.
 They are related to the $\lambda$'s as
\begin{eqnarray}
\nonumber
 \lambda_0=\cos\theta_0,~~~ 0<\theta_0<\pi,\\
\lambda_{1,2}=\cosh(\theta_1\pm\theta_2),~~~ 0<\theta_{1,2}<\infty.
\end{eqnarray}
The Jacobian is accordingly modified to
\begin{equation}
\mathcal{J}= \frac{2\sin^3\theta_0 \sinh\theta_1 \sinh\theta_2}{[\cosh(\theta_1+\theta_2)-\cos\theta_0]^2 [\cosh(\theta_1-\theta_2)-\cos\theta_0]^2}.
\end{equation}
While the numerical evaluation of the characteristic function does not pose any serious difficulties for reasonable values of $k$, it is extremely ill conditioned for the 
distribution. The calculation of the derivatives up to 4th order in the expression for the distribution are not really feasible numerically, especially of the data generated 
from a complicated 4-fold integral. Even some noise present in the data gets amplified due to derivatives, which kills the sought after result altogether. We therefore
determine the distributions in this case with the help of Eq.~\eqref{Ps2}, considering a cut-off for $k$. This approach works well for a sufficiently flat distribution, 
whereas, if it is highly localized, it is advantageous to consider the corresponding characteristic function instead. The numerical simulations were performed using 
random matrices similar to the $\beta=2$ case with the matrices now drawn from the GOE. In Fig.~\ref{Fig4} we show the comparison between the analytical predictions
and simulation results. The distributions have been obtained by taking the Fourier transform of the characteristic function numerically, as described above. We find 
perfect agreement in all cases.  

\begin{figure}[h]
\centering
\includegraphics*[width=0.7\textwidth]{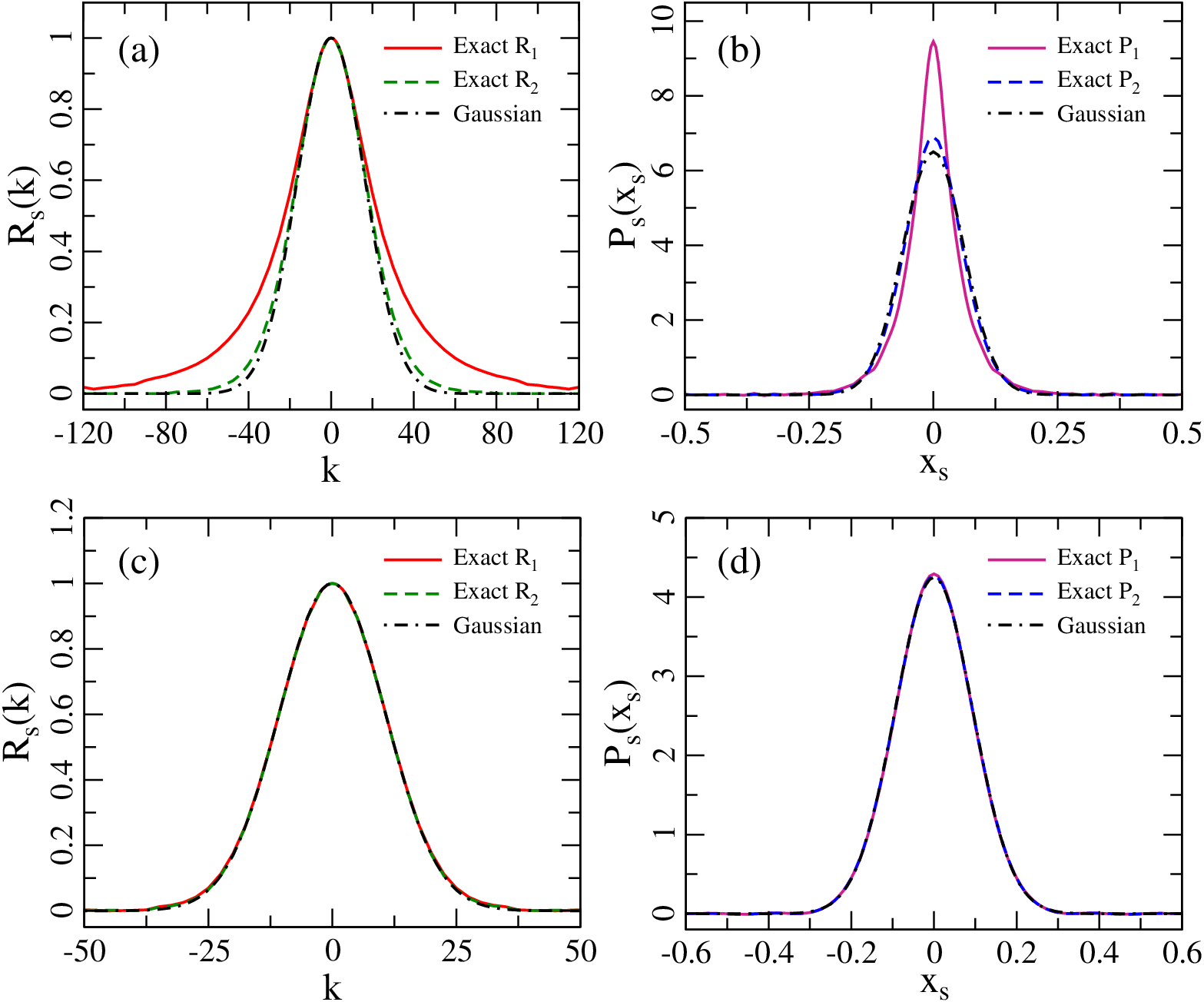}
\caption[]{Test of the approximate results in Eqs.~\eqref{Psapprox} and \eqref{Rsapprox} for the element $S_{12}$ in the $\beta=1$ case. Figs. (a), (b) show the characteristic 
functions and distributions for $\Gamma/\Delta_m\approx1.273$. Figs. (c), (d) show the same quantities for $\Gamma/\Delta_m\approx7.162$. The solid lines correspond
to the exact results for the real part, dashed lines correspond to exact results for the imaginary part, and dot-dashed lines represent the Gaussian approximations. 
The Gaussian approximations work nicely in the regime closer to Ericson.}
\label{Fig5}
\end{figure}

Since no odd powers appears when the right hand side of Eq.~\eqref{RkOE} is series-expanded in powers of $k$, it is clear that all the odd moments for both the real and
imaginary parts of off-diagonal matrix elements are zero. The expression for the second moment can be obtained by examining the coefficient of $k^2$. We have
\begin{eqnarray}
\label{beta1sec}
\nonumber
\overline{x_s^2}\!\!&=&\!\!\!4\int_{-1}^1 d\lambda_0 \int_1^\infty d\lambda_1\int_1^\infty d\lambda_2 ~\mathcal{J}(\lambda_0,\lambda_1,\lambda_2)\mathcal{F}_\text{O}(\lambda_0,\lambda_1,\lambda_2)\left(4p_a^0p_b^0+p_a^{+}p_b^{+}+p_a^{-}p_b^{-}\right)\\
&=&\!\!\!8\int_{-1}^1 d\lambda_0 \int_1^\infty d\lambda_1\int_1^\infty d\lambda_2 ~\mathcal{J}(\lambda_0,\lambda_1,\lambda_2)\mathcal{F}_\text{O}(\lambda_0,\lambda_1,\lambda_2)\left(2p_a^0p_b^0+p_a^1p_b^1+p_a^2p_b^2\right),
\end{eqnarray}
which is in complete agreement with the earlier result~\cite{VWZ1985}. We note that the above expression contains $p_c^j$ only, and therefore implies identical
 second moments for the real and imaginary parts. The fourth and higher even moments depend on $q_c^{\pm}$ and $r_c^{\pm}$ also, and therefore are different
  for the real and imaginary parts. The fourth moment is obtained by identifying the coefficient of $k^4$ in Eq.~\eqref{RkOE} as
\begin{eqnarray}
\label{beta1for}
\nonumber
\overline{x_s^4}=48\int_{-1}^1 d\lambda_0 \int_1^\infty d\lambda_1\int_1^\infty d\lambda_2 ~\mathcal{J}(\lambda_0,\lambda_1,\lambda_2)\mathcal{F}
(\lambda_0,\lambda_1,\lambda_2)\big[(p_a^{+}p_b^{+}+p_a^{-} p_b^{-})(q_a^{-}q_b^{-}+r_a^{-}r_b^{-}-2q_a^{+}q_b^{+}-2r_a^{+}r_b^{+})\\
\nonumber
-(p_a^{+}p_b^{-}+p_a^{-}p_b^{+})(q_a^{+}q_b^{-}+q_a^{-}q_b^{+}+r_a^{+}r_b^{-}+r_a^{-}r_b^{+})+4p_a^0 p_b^0(q_a^{-} q_b^{-}+r_a^{-} r_b^{-})\\
+2(4p_a^0 p_a^0+p_a^{+} p_a^{+}+p_a^{-}p_a^{-}+q_a^{+}r_a^{+})(4p_b^0 p_b^0+p_b^{+}p_b^{+}+p_b^{-} p_b^{-}+q_b^{+}r_b^{+})
\big].
\end{eqnarray}
The explicit expressions for the sixth and higher moments get extremely lengthy and therefore we refrain from presenting them here. The Mathematica codes for computation of 
the characteristic functions and moments can be found as the supplemental material available with Ref.~\cite{NKSG2014}.

In Fig.~\ref{Fig5} we consider the comparison of the exact results with Gaussian approximations as given in Eqs.~\eqref{Psapprox} and \eqref{Rsapprox}, with 
the variances $\mathcal{V}^2$ now determined by Eq.~\eqref{beta1sec}. In Figs.~\ref{Fig5} (a) and (b) we show the characteristic functions and distributions for $M=32$ 
with $T_c=0.25\,(g_c^{+}=7, g_c^{-}=-6.928)$ for all channels. Similarly in Figs.~\ref{Fig5} (c), (d) we have these quantities for $M=50$ with identical value of 
$T_c=0.9\,(g_c^{+}=1.222, g_c^{-}=-0.703)$. The Weisskopf estimate, Eq.~\eqref{Weisskopf}, gives $\Gamma/\Delta_m\approx1.273$ for the former case and we
can see deviations from the Gaussian results. In the latter case we have $\Gamma/\Delta_m\approx7.162$, with the Gaussian approximations working extremely well. 

\section{Conclusion}
\label{secConc}

We presented a detailed derivation of the full distributions for the real and imaginary parts of the off-diagonal $S$-matrix elements. These results are completely general 
and applicable to all regimes, ranging from isolated resonances to strongly overlapping resonances. Our derivation is based on the Heidelberg approach, which is the
 most general formulation of any scattering process in which an interaction region and scattering channels can be identified. To accomplish this task we introduced a
 novel route to the sigma model based on the characteristic function and thereby developed an important extension of the Supersymmetry method which led to the 
 solution of a problem which could not be tackled by the previous variants of the Supersymmetry method or any other technique. We believe that the present formalism
 will also find applications in other problems.

We verified our analytical results by numerical simulations and found excellent agreements. In Ref.~\cite{KNSGDMRS2013} we also compared our $\beta=1$ results
 with experimental data obtained from microwave experiments and thereby confirmed, for the first time, universality in the case of the off-diagonal matrix elements. 
 The generality of our results makes it possible to carry out additional investigations for other experiments and test universality there. Adding to this point we would also
 like to underline the relevance of our results beyond Schr\"odinger or Schr\"odinger--like wave dynamics.  While the dynamics of the waves in flat microwave cavities 
 mathematically coincides with the time-independent Schr\"odinger equation, it is quite different for genuine {\it classical} wave systems, such as three-dimensional
microwave cavities and vibrating elastic solids. Our results on the distribution facilitate more detailed tests than previously feasible concerning the question whether
 the fluctuations in these ``non--Schr\"odinger" systems are described by a random matrix ansatz or not.

 For the $\beta=2$ case we found that the phases of the $S$-matrix elements are uniformly distributed and hence were able to calculate an exact distribution for the
 moduli as well. The distribution of moduli is important from the point of view of experiments where only cross-sections are available. For $\beta=1$ the derivation of
 the distributions of the phases and moduli remains an unsolved task. 

\section*{Acknowledgments}
This work was supported by the Collaborative Research Center SFB/TR12, ``Symmetries and Universality in Mesoscopic Systems'' of the Deutsche Forschungsgemeinschaft.



\appendix

\section{Integral over supervectors}
\label{AppSintegral}


We outline here the important steps implemented in carrying out the integral over the supervectors. We consider the $\beta=2$ case first. Let us introduce
\begin{equation}
\widetilde{\bSg}=\mathbf{K}^{1/2} \bSg \mathbf{K}^{1/2}=\begin{bmatrix} \widetilde{\bSg}_{11} & \widetilde{\bSg}_{12} \\ \widetilde{\bSg}_{21} & \widetilde{\bSg}_{22} 
\end{bmatrix}.
\label{tildeSigma}
\end{equation}
The $\Psi$-integral can therefore be written in terms of $z$ and $\zeta$ vectors as
\begin{equation}
\int\! \dd[z] \exp\left(\frac{i}{2}(z^\dag W+U_s^\dag z)+iz^\dag \widetilde{\bSg}_{11}z \right) 
   \int\! \dd[\zeta] \exp\left(iz^\dag \widetilde{\bSg}_{12} \zeta + i\zeta^\dag \widetilde{\bSg}_{21}z + i\zeta^\dag \widetilde{\bSg}_{22} \zeta \right).
\end{equation}
Here we introduced $U_s^\dag=[-i^s W_b^\dag,(-i)^sW_a^\dag]$, and $\mathbf{U}_s^\dag$ defined below Eq.~\eqref{RkU} is $[U_s^\dag,0]$.
As we can see now, the integrals which have to be performed are just Gaussian ones. 
With equations \eqref{gintc} and \eqref{gintac} the result is
\begin{equation}
\label{afterpsi}
\det \left(\frac{\widetilde{\bSg}_{22}}{\widetilde{\bSg}_{11} - \widetilde{\bSg}_{12}\widetilde{\bSg}_{22}^{-1}\widetilde{\bSg}_{21}} \right) 
\exp\left[-\frac{i}{4}U_s^\dag \left(\widetilde{\bSg}_{11} - \widetilde{\bSg}_{12}\widetilde{\bSg}_{22}^{-1}\widetilde{\bSg}_{21}\right)^{-1}W\right].
\end{equation}
The determinant in the above equation is the inverse of the 
superdeterminant of $\mathbf{K}^{1/2} \Sigma \mathbf{K}^{1/2}$. Furthermore using the multiplicative property, we have
\begin{equation}
\sdet \mathbf{K}^{1/2}\bSg \mathbf{K}^{1/2}=\sdet \mathbf{K}\,\sdet \bSg=(-1)^N \sdet \bSg.
\end{equation}
The last step follows since $\sdet\mathbf{K}=(-1)^N$.  The factor of $(-1)^N$ generated above cancels the same factor in Eq.~\eqref{susym} leading to Eq.~\eqref{Rksigma}. 
The matrix in the exponential in Eq. \eqref{afterpsi} can be identified as the upper left block of the inverse of $\mathbf{K}^{1/2} \bSg \mathbf{K}^{1/2}$. Thus we get
\begin{equation}
U_s^\dag \left(\widetilde{\bSg}_{11} - \widetilde{\bSg}_{12}\widetilde{\bSg}_{22}^{-1}\widetilde{\bSg}_{21}\right)^{-1} W 
=U_s^\dag \Big(\left(\mathbf{K}^{1/2} \bSg \mathbf{K}^{1/2} \right)^{-1} \Big)_{11} W 
=\mathbf{U}_s^\dag \left(\mathbf{K}^{1/2} \bSg \mathbf{K}^{1/2} \right)^{-1} \mathbf{W}.
\end{equation}
The second equality above follows since the vectors $\mathbf{U}_s$ and $\mathbf{W}$ have zeros in their lower halves. For the same reason we
 may replace $\mathbf{K}$ in the above equation by $\mathbf{L}$ which differ only in their lower halves. Taking everything into account we arrive at Eq. \eqref{Rksigma}.


We now consider $\beta=1$. The supervector $\Psi$ in this case has a more complicated structure and we cannot use the results from the $\beta=2$ calculation. 
Since the commuting part $z$ of $\Psi$ is now real and the anticommuting part $\zeta$ comprises both the anticommuting variables and their complex conjugates 
we cannot  use the identities for Gaussian integrals, Eqs. \eqref{gintc}, \eqref{gintac}. However, for a real vector $z$ we have a similar identity:
\begin{equation}
 \int\! \dd[z] \exp\left(i z^T \mathbf{a} \,z + i z^T \mathbf{b} \right)=\sqrt{\det\left(\frac{\pi i}{\mathbf{a}}\right)} \exp\left(-\frac{i}{4} \mathbf{b}^T \mathbf{a}^{-1} \mathbf{b} \right), 
 \quad \mathbf{a}^T=\mathbf{a}.
 \label{gauss_real}
\end{equation}
The main difference to the Gaussian integral over the complex vector is the appearance of the square root over the determinant. As indicated in the above equation, 
the matrix $\mathbf{a}$ has to be symmetric in this case.
A similar identity holds for a Grassmann vector also, viz.
\begin{equation}
  \int\! \dd[\zeta] \exp\left(i\zeta^T \mathbf{a} \zeta\right)= \sqrt{\det\left(\frac{\mathbf{a}}{i \pi} \right)}, \quad \mathbf{a}^T=-\mathbf{a}.
  \label{gauss_areal}
\end{equation}
However, we note that $\mathbf{a}$ is skew-symmetric in this case. One could also derive a more general result adding a linear term $\zeta^T \mu$ in the exponent. 
This would give rise to an additional exponential factor besides the determinant. For our purposes the above identity suffices.

Similar to the $\beta=2$ case above, we again define $\widetilde{\bSg}$. In order to use the identities \eqref{gauss_real} and \eqref{gauss_areal}, we have to look
at an expression involving $\Psi^T=[z^T,\ \zeta^T]$ rather than $\Psi^\dag=[z^\dag,\ \zeta^\dag]$. For this we observe that 
$\Psi^\dag$ and $\Psi^T$ are related via a projection matrix $J$ as
\begin{equation}
  \Psi^\dag = \Psi^T J,
\end{equation}
where
\begin{equation}
  J=\begin{bmatrix} \1_{4N} & 0 \\ 0 & J_2 \end{bmatrix}, ~~~~J_2= \begin{bmatrix} 0 & -\1_N & 0 & 0 \\ +\1_N & 0 & 0 & 0 \\ 0 & 0& 0 & -\1_N \\ 0 & 0 & +\1_N & 0 \end{bmatrix}.
\end{equation}
Note that $J_2$ is orthogonal and skew-symmetric, $J_2^{-1}=J_2^T=-J_2$.
Instead of $\Psi^\dag \widetilde{\bSg} \Psi$ we therefore look at 
\begin{equation}
  \Psi^T J \widetilde{\bSg} \Psi = z^T \widetilde{\bSg}_{11} z + z^T \widetilde{\bSg}_{12} \zeta + \zeta^T J_2 \widetilde{\bSg}_{21} z + \zeta^T J_2 \widetilde{\bSg}_{22} \zeta.
 \label{bfGOE}
\end{equation}
It is not difficult to verify that $\widetilde{\bSg}_{11}$ is symmetric,
\begin{equation}
 \label{Sgm11}
 \widetilde{\bSg}_{11}^T=\widetilde{\bSg}_{11},
\end{equation}
 while $J_2 \widetilde{\bSg}_{22}$ is skew-symmetric,
\begin{equation}
\label{Sgm22}
 (J_2 \widetilde{\bSg}_{22})^T=-J_2 \widetilde{\bSg}_{22}.
\end{equation}
We also
 observe that the off-diagonal blocks are connected via
\begin{equation}
\label{Sgm21}
  J_2 \widetilde{\bSg}_{21} = - \widetilde{\bSg}_{12}^T.
\end{equation}
Thus the off-diagonal blocks of $J \widetilde{\bSg}$ are the negative transpose of each other.

In Eq.~\eqref{bfGOE} the commuting and anticommuting variables are mixed, but in order to use the integral identities \eqref{gauss_real}
 and \eqref{gauss_areal} it would be more convenient if they were separated. We can achieve this by carrying out the transformation
\begin{equation}
  \zeta \rightarrow \zeta - \widetilde{\bSg}_{22}^{-1}\widetilde{\bSg}_{21} z.
\end{equation}
Since $\zeta^T$ is not independent of $\zeta$, this implies further that
\begin{equation}
  \zeta^T \rightarrow \zeta^T - z^T\widetilde{\bSg}_{21}^T (\widetilde{\bSg}_{22}^{-1})^T
 = \zeta^T - z^T\widetilde{\bSg}_{12} \widetilde{\bSg}_{22}^{-1} J_2^{-1},
\end{equation}
where we used Eqs. (\ref{Sgm22}) and (\ref{Sgm21}). This transformation does not change the value of the integral.
The bilinear forms in \eqref{bfGOE} change accordingly,
\begin{equation}
\nonumber
 \Psi^T J \widetilde{\bSg} \Psi \rightarrow z^T (\widetilde{\bSg}_{11}- \widetilde{\bSg}_{12}\widetilde{\bSg}_{22}^{-1}\widetilde{\bSg}_{21}) z + \zeta^T J_2 \widetilde{\bSg}_{22} \zeta,
\end{equation}
and therefore the $\Psi$-integral becomes
\begin{equation}
  \int\! \dd[z] \exp\left(i\, z^T V_s + i z^T (\widetilde{\bSg}_{11}- \widetilde{\bSg}_{12}\widetilde{\bSg}_{22}^{-1}\widetilde{\bSg}_{21}) z \right)
  \int\! \dd[\zeta] \exp\left(i \zeta^T J_2 \widetilde{\bSg}_{22} \zeta \right).
\end{equation}
We have used here $\mathbf{V}_s^T=[V_s^T,0]$.
Since $J_2 \widetilde{\bSg}_{22}$ is skew-symmetric we can use Eq.~\eqref{gauss_areal} to evaluate the integral over the anticommuting
 variables and get as result
\begin{equation}
  \int\! \dd[\zeta] \exp\left(i \zeta^T J_2 \widetilde{\bSg}_{22} \zeta \right) = {\det}^{1/2} \left(\frac{J_{2}\widetilde{\bSg}_{22}}{i \pi} \right) = {\det}^{1/2} 
  \left(\frac{\widetilde{\bSg}_{22}}{i \pi} \right),
\end{equation}
where we used that $\det J_2 = 1$.

With the properties \eqref{Sgm11}, \eqref{Sgm22} and \eqref{Sgm21} it is easy to check that $\tilde{\Sigma}_{11}-\tilde{\Sigma}_{12} \tilde{\Sigma}_{22} \tilde{\Sigma}_{21}$ is symmetric and hence Eq. \eqref{gauss_real} is applicable.
Altogether the integration over the supervector yields
\begin{equation}
  {\det}^{1/2} \left(\frac{\widetilde{\bSg}_{22}}{\widetilde{\bSg}_{11}-\widetilde{\bSg}_{12} \widetilde{\bSg}_{22}^{-1} \widetilde{\bSg}_{21}} \right)
  \exp \left( -\frac{i}{4} V_s^T \left(\widetilde{\bSg}_{11}-\widetilde{\bSg}_{12} \widetilde{\bSg}_{22}^{-1} \widetilde{\bSg}_{21} \right)^{-1} V_s \right).
\end{equation}
This leads to Eq.~\eqref{RksigmaOE} using arguments similar to that in the $\beta=2$ case.


\section{Derivation of Eqs.~\eqref{Lnsdet} and~\eqref{Sgminv}}
\label{AppET}

We give in this appendix the proofs for Eqs.~\eqref{Lnsdet} and~\eqref{Sgminv}. The steps below apply to the $\beta=1$ case also 
and thus we use the notations $\1_{8/\beta}$ instead of $\1_4$ etc. 
The superdeterminant part can be split into two parts,
\begin{eqnarray}
\label{strlnSig}
\text{str} \ln {\boldsymbol\Sigma} \!\!\!&=&\!\!\! \str \ln \left[ \sigma_E \otimes \1_N 
+\frac{i}{4k} L \otimes \left( \sum_{c=1}^M W_c W_c^\dag \right) \right] \nonumber \\ \label{strlnSigma}
                &=&\!\!\! N\str \ln \sigma_E
                +\str \ln \left[\1_{8N/\beta}+\frac{i}{4k}\sigma_E^{-1} L
                       \otimes \left( \sum_{c=1}^M  W_c W_c^\dag \right) \right].
\end{eqnarray}
The second term of \eqref{strlnSigma} can be cast into the form 
\begin{equation}
\str \ln \left[\1_{8N/\beta}+\frac{i}{4k} \sigma_E^{-1} L
                       \otimes \left( \sum_{c=1}^M  W_c W_c^\dag \right) \right]
=\sum_{c=1}^M \text{str} \ln \left(\mathds{1}_{8/\beta} + \frac{i\gamma_c}{4\pi k}\sigma_E^{-1}L \right).
\end{equation}
Therefore one needs to expand the logarithm into a Taylor series, employ the orthogonality relation, Eq. \eqref{Wortho}, of the coupling vectors and rewrite the series again into a logarithm. 
These results lead to Eq.~\eqref{Lnsdet}.

We now outline the steps to invert $ {\boldsymbol\Sigma}$. The main idea in the following calculation is to write the inverse as a series expansion and then employ the orthogonality relation, Eq. (6), as done above to arrive at the Eq. (B.2).

\begin{eqnarray}
 {\boldsymbol\Sigma}^{-1}\!\!\!&=&\!\!\!\left( \sigma_E \otimes \1_N +\frac{i}{4k}L \otimes \sum_{c=1}^M W_c W_c^\dag \right)^{-1} \nonumber \\
 &= &\!\!\!\left( \sigma_E^{-1} \otimes \1_N \right) \left[\1_{8N/\beta} +\sum_{n=1}^\infty \left(-\frac{i}{4k} L \sigma_E^{-1}\right)^n \otimes  \sum_{c=1}^M \left(\frac{\gamma_c}{\pi}\right)^{n-1} W_c W_c^\dag \right].
\end{eqnarray}
To rewrite the series into an inverse again, we have to take care that the term for
 $n=0$ is missing (note the additional $-\1_{8/\beta}$ in the next line). Then we get
\begin{eqnarray}
 {\boldsymbol\Sigma}^{-1}\!\!\!&=&\!\!\!\left( \sigma_E^{-1} \otimes \1_N \right) \left[\1_{8N/\beta} +\sum_{c=1}^M \left( \left(\1_{8/\beta}
 +\frac{i \gamma_c}{4\pi k} L \sigma_E^{-1}\right)^{-1} -\1_{8/\beta} \right) \otimes  \frac{\pi}{\gamma_c}W_c W_c^\dag \right] \nonumber \\
 &=&\!\!\!\sigma_E^{-1} \otimes \1_N - \sum_{c=1}^M \sigma_E^{-1} \otimes \frac{\pi}{\gamma_c}W_c W_c^\dag+ \sum_{c=1}^M \left(\sigma_E 
 + \frac{i \gamma_c}{4 \pi k} L \right)^{-1} \otimes \frac{\pi}{\gamma_c}W_c W_c^\dag,
\label{invSig}
\end{eqnarray}
which eventually gives Eq.~\eqref{Sgminv}.


\section{Derivation of Eq.~\eqref{VLSLW}}
\label{AppVLSLW}

We use Eq.~\eqref{Sgminv} to calculate $\mathbf{U}_s^\dag \mathbf{L}^{-1/2}  {\boldsymbol\Sigma}^{-1} \mathbf{L}^{-1/2} \mathbf{W}$. The first term in 
Eq.~\eqref{Sgminv} yields  $\mathbf{U}_s^\dag \mathbf{L}^{-1/2} (\sigma_E^{-1} \otimes \1_N) \mathbf{L}^{-1/2} \mathbf{W}$. Recalling the definitions for 
$\mathbf{U}_s$ and $\mathbf{W}$ we obtain the following bilinear form:
\begin{eqnarray}
  \label{UxV1}
\nonumber
\mathbf{U}_s^\dag \mathbf{L}^{-1/2} (\sigma_E^{-1} \otimes \1_N) \mathbf{L}^{-1/2} \mathbf{W}
 =\begin{bmatrix} -i^s W_b^\dag & (-i)^s W_a^\dag \end{bmatrix}
\begin{bmatrix}
  (\sigma_E^{-1})_{11} \1_N & -i(\sigma_E^{-1})_{12} \1_N \\
  -i(\sigma_E^{-1})_{21} \1_N & -(\sigma_E^{-1})_{22} \1_N
\end{bmatrix}
 \begin{bmatrix} W_a \\ W_b \end{bmatrix}\\
  =-i^s (\sigma_E^{-1})_{11} W_b^\dag W_a +i^{s+1}(\sigma_E^{-1})_{12} W_b^\dag W_b  
 +(-i)^{s+1}(\sigma_E^{-1})_{21} W_a^\dag W_a -(-i)^s (\sigma_E^{-1})_{22} W_a^\dag W_b.
\end{eqnarray}
Since we are interested in the off-diagonal elements of the scattering matrix, i.e. $a \neq b$, owing to the orthogonality relation, Eq.  \eqref{Wortho}, 
this implies that the first and the last term in Eq.~\eqref{UxV1} vanish, while the two other terms are given by
\begin{equation}
\label{UxV2}
 \frac{i^{s+1}\gamma_b}{\pi}(\sigma_E^{-1})_{12} + \frac{(-i)^{s+1}\gamma_a}{\pi}(\sigma_E^{-1})_{21}. 
\end{equation}
The bilinear form of the second term of Eq.~\eqref{Sgminv} can be calculated in a similar way and yields exactly the same result as \eqref{UxV2}.
 Since these two terms appear with opposite signs in Eq. \eqref{Sgminv} they cancel out and we are left with only the last term,
\begin{equation}
\mathbf{U}_s^\dag \mathbf{L}^{-1/2} \left(\sum_{c=1}^M \left(\sigma_E - \frac{i \gamma_c}{4 \pi k} L \right)^{-1} \otimes \frac{\pi}{\gamma_c}
W_c W_c^\dag\right)\mathbf{L}^{-1/2} \mathbf{W}.
  \label{Sigmainv}
\end{equation}
It has the same structure as the second term in Eq. \eqref{Sgminv} and therefore we get for the second exponential factor in Eq. \eqref{Rksigma},
\begin{equation}
\mathbf{U}_s^\dag \left(\mathbf{L}^{1/2} {\boldsymbol\Sigma} \mathbf{L}^{1/2} \right)^{-1} \mathbf{W} 
 = \frac{i^{s+1}\gamma_b}{\pi}\rho^{(b)}_{12} +\frac{(-i)^{s+1}\gamma_a}{\pi}\rho^{(a)}_{21},
 \label{expoGUE}
\end{equation}
with $\rho^{(c)}$ as defined in Eq.~\eqref{rho}.


\section{Parametrization of the supermatrices}
\label{AppParam}

The $4\times 4$ and $8\times 8$ supermatrices used in the calculations for the $\beta=1$ and 2 cases can be parametrized as
\begin{equation}
\label{Qparam}
 Q=\mathcal{U}^{-1} \mathcal{D}\, \mathcal{U},
\end{equation}
where $\mathcal{U}$ is a unitary supermatrix and $\mathcal{D}$ is a quasidiagonal matrix. These matrices have the following block structure in [1,2] (or $pq$) 
notation~\cite{VWZ1985,FS1997}
\begin{equation}
\label{Block}
\mathcal{U}=\begin{bmatrix}
		u_1 & 0 \\
		0 & u_2 
	      \end{bmatrix},~~~~~~~
\mathcal{D}=\begin{bmatrix}
	      -i \mathcal{D}_1 & \mathcal{D}_{12} \\
	       \mathcal{D}_{21} & i \mathcal{D}_1
	     \end{bmatrix}.
\end{equation}
For the $4\times4$ supermatrix we have \cite{FS1997}
\begin{equation}
\mathcal{D}_1=\begin{bmatrix}
     \lambda_1 & 0 \\
      0 & \lambda_2
    \end{bmatrix},~~~~~~
\mathcal{D}_{12}=\begin{bmatrix}
     (\lambda_1^2-1)^{1/2} e^{i \phi_1} & 0 \\
      0 & i(1-\lambda_2^2)^{1/2} e^{-i \phi_2}
    \end{bmatrix},~~~~~~
\mathcal{D}_{21}=\begin{bmatrix}
     (\lambda_1^2-1)^{1/2} e^{-i \phi_1} & 0 \\
      0 & i(1-\lambda_2^2)^{1/2}e^{i \phi_2}
    \end{bmatrix},
\end{equation}
\begin{equation}
u_1=\begin{bmatrix}
   1-\frac{1}{2}\alpha^* \alpha & -\alpha^* \\
   \alpha &  1+\frac{1}{2}\alpha^* \alpha
  \end{bmatrix},~~~~~~~~~
u_2=\begin{bmatrix}
   1+\frac{1}{2}\beta^* \beta & -i\beta^* \\
   i\beta &  1-\frac{1}{2}\beta^* \beta
  \end{bmatrix}.
\end{equation}
Here $\alpha, \alpha^*, \beta, \beta^*$ are Grassmann variables, $\lambda_1 \in (1,\infty),\lambda_2 \in (-1,1)$, and $ \phi_1, \phi_2 \in (0,2\pi)$. 
Also, $u_1^{-1}=u^\dagger$, $u_2^{-1}=\boldsymbol{l} u_2^\dagger \boldsymbol{l}$ with $\boldsymbol{l}=\text{diag}(1,-1)$. The corresponding measure is given by
\begin{equation}
d Q=\frac{d\lambda_1 d\lambda_2}{(\lambda_1-\lambda_2)^2}\frac{d\phi_1 d\phi_2}{(2 \pi)^2}d\alpha^* d\beta^* d\alpha d\beta.
\end{equation}

For the 8$\times$8 supermatrix  $Q$, we take the parametrization from~\cite{VWZ1985}. $\mathcal{D}$ is parametrized in terms of the pseudo eigenvalues
 $\lambda_j, j=0,1,2$, and SU(2) variables $m,r,s$. We have,
\begin{eqnarray}
\nonumber
\mathcal{D}_1=\text{diag}(\lambda_1,\lambda_2,\lambda_0,\lambda_0),\hspace{6cm}\\
\mathcal{D}_{12}=\begin{bmatrix}
	(\lambda_1^2-1)^{1/2} & 	& \\
	      & (\lambda_2^2-1)^{1/2} 	& \\
	      &		& i(1-\lambda_0^2)^{1/2} U
    \end{bmatrix},~~~
\mathcal{D}_{21}=\begin{bmatrix}
	(\lambda_1^2-1)^{1/2} & 	& \\
	      & (\lambda_2^2-1)^{1/2} 	& \\
	      &		& i(1-\lambda_0^2)^{1/2} U^\dagger
    \end{bmatrix},
\end{eqnarray}
with
\begin{equation}
U=\frac{1}{\sqrt{1+m^2+r^2+s^2}} \begin{bmatrix}
                                1+ im & -r-is \\
				r-is & 1-im
                               \end{bmatrix}.
\end{equation}
Here $\lambda_0 \in (-1,1), \lambda_{1,2} \in (1,\infty)$ and $m,r,s \in (-\infty,\infty)$. We note that $UU^\dag=U^\dag U=\mathds{1}_2$.
The $\lambda$'s used here are different from those in \cite{VWZ1985} which we denote by $\widetilde{\lambda}$. They are related as
$\widetilde{\lambda}_{1,2}^2=\lambda_{1,2}^2-1,\widetilde{\lambda}_0^2=1-\lambda_0^2$. The Jacobian measure has been changed accordingly.

The blocks of $\mathcal{U}$ are given in terms of O(2) variables $\phi_1, \phi_2$ and 8 Grassmann variables as,
\begin{equation}
u_i=O_i v_i,~~~
O_i=\begin{bmatrix}
     \cos \phi_i & \sin \phi_i & 0 & 0 \\
     -\sin \phi_i & \cos \phi_i & 0 & 0 \\
      0 & 0 & 1 & 0 \\
      0 & 0 & 0 & 1
    \end{bmatrix}, ~~ i=1,2. 
\end{equation}
The $v$'s are defined using 
\begin{equation}
Y_i=\begin{bmatrix}
      0 & 0 & -\alpha_i^* & \alpha_i \\
      0 & 0 & -\beta_i^* & \beta_i \\
      \alpha_i & \beta_i & 0 & 0 \\
      \alpha_i^* & \beta_i^* & 0 & 0
    \end{bmatrix}
\end{equation}
as
\begin{eqnarray}
\nonumber
v_1^{\pm1}=1\pm Y_1+\tfrac{1}{2}Y_1^2 \pm \tfrac{1}{2}Y_1^3+\tfrac{3}{8}Y_1^4~~~\\
v_2^{\pm1}=1\pm iY_2-\tfrac{1}{2}Y_2^2 \mp \tfrac{1}{2}iY_2^3+\tfrac{3}{8}Y_2^4.
\end{eqnarray}
The measure $d\mu(Q)$ is given by
\begin{equation}
d\mu(Q)=d\mu(\lambda_0,\lambda_1,\lambda_2) d\mu(\text{SU}(2)) d\mu(\text{O}(2)) d\mu_{Gr},
\end{equation}
with
\begin{eqnarray}
\nonumber
d\mu(\lambda_0,\lambda_1,\lambda_2)=\frac{1}{2}\frac{(1-\lambda_0^2)|\lambda_1-\lambda_2|}
{(\lambda_1^2-1)^{1/2}(\lambda_2^2-1)^{1/2}(\lambda_1-\lambda_0)^2 (\lambda_2-\lambda_0)^2}d\lambda_0 d\lambda_1 d\lambda_2 ,\\
\nonumber
d\mu(\text{SU}(2))=\frac{1}{(1+m^2+r^2+s^2)^2} dm dr ds,~~~d\mu(\text{O}(2))=d\phi_1 d\phi_2,\\
d\mu_{Gr}=d\alpha_1 d\alpha_1^* d\alpha_2 d\alpha_2^* d\beta_1 d\beta_1^* d\beta_2 d\beta_2^*.~~~~~~~~~~~~~~~~~~~~~~
\end{eqnarray}


\section{Calculation of $\rho^{(c)}$}
\label{AppRho}

Using Eq.~\eqref{rho}, we have
\begin{equation}
 \rho^{(c)} 
= \left(\1_{8/\beta} +\frac{i \gamma_c}{4 \pi k} \sigma_E^{-1} L \right)^{-1} \sigma_E^{-1}
= -\frac{16 \pi^2 k^2}{v^2} \left(\1_{8/\beta} -\frac{i 4 \pi k \gamma_c}{v^2} \sigma_{G} L \right)^{-1} \sigma_{G},
\end{equation}
where in the second step above we have used the saddle point condition, Eq. \eqref{SPeq}.
Using Eq. \eqref{SPmnfld} for $\sigma_G$ and $Q=\mathcal{U}^{-1}\mathcal{D} \mathcal{U}$ we get
\begin{equation}
 \rho^{(c)}=-\frac{2 \pi k}{v^2} \mathcal{U}^{-1 }
\left(\1_{8/\beta} -\frac{i \gamma_c E}{2 v^2}  L+\frac{i \gamma_c \Delta}{2 v^2} \mathcal{D} L \right)^{-1} \mathcal{U} 
(E\1_{8/\beta} - \Delta Q),
 \label{rhointermediate}
\end{equation}
where we have used that $\mathcal{U}$ and $L$ commute. Thus we need to calculate $\mathcal{R}^{-1}$ where
\begin{equation}
\mathcal{R}= \left(\1_{8/\beta} -\frac{i \gamma_c E}{2 v^2} L+\frac{i \gamma_c \Delta}{2 v^2} \mathcal{D} L \right).
\end{equation}
We propose the ansatz
\begin{equation}
\mathcal{R}^{-1}=\mathbf{d} \left(\1_{8/\beta}+\frac{i \gamma_c E}{2v^2}L+\frac{i \gamma_c \Delta}{2v^2}L \mathcal{D}\right),
\end{equation}
where {\bf d} is a diagonal matrix to be determined. Using the results $\mathcal{D}^2=-\1_{8/\beta}$ and $L^2=\1_{8/\beta}$,
 we obtain from
\begin{equation}
\1_{8/\beta}=\mathcal{R}^{-1} \mathcal{R}=\mathbf{d} \left(\1_{8/\beta}+\frac{i \gamma_c E}{2v^2}L+\frac{i \gamma_c \Delta}{2v^2}L \mathcal{D}\right)
\left(\1_{8/\beta}-\frac{i \gamma_c E}{2v^2}L+\frac{i \gamma_c \Delta}{2v^2}\mathcal{D}L\right),  
\end{equation}
\begin{equation}
 \mathbf{d}^{-1}=\left(1+\frac{\gamma_c^2}{v^2}\right)\1_{8/\beta}+\frac{i\gamma_c \Delta}{2v^2}(\mathcal{D}L+L\mathcal{D}).
\end{equation}
With $\mathcal{D}$ given in Eq.~\eqref{Block}, the above equation eventually gives in [1,2] (or $pq$) notation,
\begin{eqnarray}
 \mathbf{d}^{-1}=\begin{cases}
             (\gamma_c\Delta/v^2)\,\text{diag}(g_c^{+}+\lambda_1,g_c^{+}+\lambda_2,g_c^{+}+\lambda_0,g_c^{+}+\lambda_0,
	     g_c^{+}+\lambda_1,g_c^{+}+\lambda_2,g_c^{+}+\lambda_0,g_c^{+}+\lambda_0)& \beta=1,\\
             (\gamma_c\Delta/v^2)\,\text{diag}(g_c^{+}+\lambda_1,g_c^{+}+\lambda_2,g_c^{+}+\lambda_1,g_c^{+}+\lambda_2) &
                        \beta=2.
            \end{cases}
\end{eqnarray}

Thus we finally have
\begin{eqnarray}
\rho^{(c)}=-\frac{2 \pi k}{v^2} \mathcal{U}^{-1 } 
 \mathbf{d} \left(\mathds{1}_{8/\beta}+\frac{i \gamma_c E}{2v^2}L+\frac{i \gamma_c \Delta}{2v^2}L \mathcal{D}\right) \mathcal{U} \left(E\1_{8/\beta} - \Delta Q\right).
\end{eqnarray}
We can now use the parametrizations given in \ref{AppParam} to obtain explicit expressions for $\rho^{(c)}$.


\section{Proof of Eq.~\eqref{Pr}}
\label{AppMod}

For $\beta=2$ the distributions for real and imaginary parts are identical, thus as already mentioned in the main text the 
joint distribution of real and imaginary parts is of the form $P_{x,y}(\sqrt{x^2+y^2})$. The distribution of the real part of $S_{ab}$
 can be obtained from this joint distribution by integrating out the imaginary part, i.e.,
\begin{equation}
\label{PxPr}
P_x(x)=2\int_0^{\sqrt{1-x^2}} \!\!\!\!dy\, P_{x,y}\left(\sqrt{x^2+y^2}\right)=\frac{1}{\pi}\int_x^1 dr\,r (r^2-x^2)^{-1/2} P_r(r). 
\end{equation}
In the second step in the above equation we switched over to the polar coordinates and used Eq. \eqref{Pphi}.

Now let us consider $f_0(x)$ which is the integral term of Eq.~\eqref{fx},
\begin{equation}
f_0(x)=\int_1^\infty d\lambda_1\int_{-1}^1 d\lambda_2 \frac{1}{4\pi(\lambda_1-\lambda_2)^2}\,\mathcal{F}_\text{U}(\lambda_1,\lambda_2)
\big(\omega_1+\omega_2\big)\big(\omega_1-x^2\big)^{-1/2}\Theta(\omega_1-x^2).
\end{equation}
Here we introduced $\omega_j=t_a^j t_b^j, j=1,2$; see Eq. \eqref{tcj}. This gives on solving for $\lambda_1$ in terms of $\omega_1$,
\begin{equation}
 \lambda_1=\frac{\omega_1(g_a+g_b)+\sqrt{\omega_1[\omega_1(g_a-g_b)^2+4g_a g_b-4]+4}}{2(1-\omega_1)}.
\end{equation}
Thus, in terms of the integration variables $\lambda_2$ and $\omega_1$ we have 
\begin{equation}
f_0(x)=\int_{-1}^1 d\lambda_2\int_{x^2}^1 d\omega_1 (\omega_1-x^2)^{-1/2} \left(\frac{\partial\lambda_1}{\partial\omega_1}\right)
\frac{\mathcal{F}_U(\lambda_1,\lambda_2)(\omega_1+\omega_2)}{4\pi(\lambda_1-\lambda_2)^2}. 
\end{equation}
Partial integration with respect to $\omega_1$ gives
\begin{equation}
 f_0(x)=-\int_{-1}^1 d\lambda_2\int_{x^2}^1 d\omega_1 \,2 (\omega_1-x^2)^{1/2} \frac{\partial}{\partial\omega_1}
\left[\left(\frac{\partial\lambda_1}{\partial\omega_1}\right)
\frac{\mathcal{F}_U(\lambda_1,\lambda_2)(\omega_1+\omega_2)}{4\pi(\lambda_1-\lambda_2)^2}\right],
\end{equation}
the boundary terms being vanishing. Therefore we have,
\begin{equation}
\frac{\partial f_0(x)}{\partial x}=\int_{-1}^1 d\lambda_2\int_{x^2}^1 d\omega_1 \,x (\omega_1-x^2)^{-1/2} \frac{\partial}{\partial\omega_1}
\left[\left(\frac{\partial\lambda_1}{\partial\omega_1}\right)
\frac{\mathcal{F}_U(\lambda_1,\lambda_2)(\omega_1+\omega_2)}{2\pi(\lambda_1-\lambda_2)^2}\right].
\end{equation}
Another partial integration with respect to $\omega_1$ in this result yields
\begin{equation}
\frac{\partial f_0(x)}{\partial x}=-\int_{-1}^1 d\lambda_2\int_{x^2}^1 d\omega_1 \,x (\omega_1-x^2)^{1/2} \frac{\partial^2}{\partial\omega_1^2}
\left[\left(\frac{\partial\lambda_1}{\partial\omega_1}\right)
\frac{\mathcal{F}_U(\lambda_1,\lambda_2)(\omega_1+\omega_2)}{\pi(\lambda_1-\lambda_2)^2}\right],
\end{equation}
which then gives
\begin{equation}
\frac{\partial^2 f_0(x)}{\partial x^2}=-\int_{-1}^1 d\lambda_2\int_{x^2}^1 d\omega_1 \,\left[(\omega_1-x^2)^{1/2}-x^2(\omega_1-x^2)^{-1/2}\right]
 \frac{\partial^2}{\partial\omega_1^2}
\left[\left(\frac{\partial\lambda_1}{\partial\omega_1}\right)
\frac{\mathcal{F}_U(\lambda_1,\lambda_2)(\omega_1+\omega_2)}{\pi(\lambda_1-\lambda_2)^2}\right].
\end{equation}
Now observing that
\begin{equation}
 (\omega_1-x^2)^{1/2}-x^2(\omega_1-x^2)^{-1/2}=-\omega_1 (\omega_1-x^2)^{-1/2}+2(\omega_1-x^2)^{1/2}
\end{equation}
we get
\begin{eqnarray}
\nonumber
 \frac{\partial^2 f_0(x)}{\partial x^2}=\int_{-1}^1 d\lambda_2\int_{x^2}^1 d\omega_1 \,\omega_1(\omega_1-x^2)^{-1/2}
 \frac{\partial^2}{\partial\omega_1^2}
\left[\left(\frac{\partial\lambda_1}{\partial\omega_1}\right)
\frac{\mathcal{F}_U(\lambda_1,\lambda_2)(\omega_1+\omega_2)}{\pi(\lambda_1-\lambda_2)^2}\right]\\
-\int_{-1}^1 d\lambda_2\int_{x^2}^1 d\omega_1 \,2(\omega_1-x^2)^{1/2}
 \frac{\partial^2}{\partial\omega_1^2}
\left[\left(\frac{\partial\lambda_1}{\partial\omega_1}\right)
\frac{\mathcal{F}_U(\lambda_1,\lambda_2)(\omega_1+\omega_2)}{\pi(\lambda_1-\lambda_2)^2}\right].
\end{eqnarray}
A final partial integration with respect to $\omega_1$ in the second term of the above equation gives
\begin{equation}
\frac{\partial^2 f_0(x)}{\partial x^2}=\int_{-1}^1 d\lambda_2\int_{x^2}^1 d\omega_1 \,(\omega_1-x^2)^{-1/2}
 \left(\omega_1\frac{\partial^2}{\partial\omega_1^2}+\frac{\partial}{\partial\omega_1}\right)
\left[\left(\frac{\partial\lambda_1}{\partial\omega_1}\right)
\frac{\mathcal{F}_U(\lambda_1,\lambda_2)(\omega_1+\omega_2)}{\pi(\lambda_1-\lambda_2)^2}\right].
\end{equation}
This result is similar in structure to Eq. \eqref{PxPr} with $\omega_1$ being identified as $r^2$. We therefore substitute $\omega_1=r^2$ and
 do some rearrangement. This leads to (for $x>0$),
\begin{equation}
\frac{\partial^2 f_0(x)}{\partial x^2}=\int_x^1 dr \,r(r^2-x^2)^{-1/2}\frac{1}{2r}
\frac{\partial}{\partial r}\left(r\frac{\partial}{\partial r}\right)\int_{-1}^1 d\lambda_2  
\left[\left(\frac{\partial\lambda_1}{\partial r}\right)
\frac{\mathcal{F}_U(\lambda_1,\lambda_2)(\omega_1+\omega_2)}{\pi(\lambda_1-\lambda_2)^2}\right]. 
\end{equation}
Using this result in Eq.~\eqref{PxUE} and comparing the resultant expression with Eq. \eqref{PxPr} we obtain Eq. \eqref{Pr}.








\end{document}